\documentclass[aps,pra,superscriptaddress,twocolumn,showpacs,longbibliography,nofootinbib]{revtex4-1}
\PassOptionsToPackage{usenames,dvipsnames}{color}
\usepackage{amsmath,amsfonts,amssymb,amsthm,graphicx,bbm,enumerate,times,color,mathrsfs}
\usepackage{dsfont}
\usepackage{times}
\usepackage[normalem]{ulem}
\usepackage{mathtools}
\usepackage[makeroom]{cancel}

\usepackage{natbib}
\setcitestyle{square,comma,numbers}
 \usepackage{amsfonts}
\usepackage{amsmath}
 \usepackage{color}
 \usepackage{amssymb}			

\usepackage[colorlinks=true,citecolor=blue,linkcolor=blue,urlcolor=blue]{hyperref}

\usepackage{amsmath,amsfonts,amssymb}
\usepackage{mathtools}
\usepackage{simpler-wick}

\newcommand{\ket}[1]{|#1\rangle}
\newcommand{\bra}[1]{\langle #1|}
\newcommand{\braket}[2]{\langle #1|#2\rangle}

\def\oned{\mathrm{1d}}
\def\dd{\mathord{\rm d}}

\newcommand{\multInt}{\int \hspace{-1mm}...\hspace{-1mm} \int}

\begin{document}

\title{Multimode Fock states with large photon number: effective descriptions and applications in quantum metrology}%
\date{\today}

\author{M.~Perarnau-Llobet}
\email{marti.perarnau@mpq.mpg.de}
\affiliation{Max-Planck-Institut f\"ur Quantenoptik, Hans-Kopfermann-Str.~1, D-85748 Garching, Germany}
\affiliation{Munich Center for Quantum Science and Technology (MCQST), Schellingstr. 4, D-80799 M\"unchen}
\author{A.~Gonz\' alez-Tudela}
\affiliation{Instituto de F\'isica Fundamental IFF-CSIC, Calle Serrano 113b, Madrid 28006, Spain.}
\author{J. I.~Cirac}
\affiliation{Max-Planck-Institut f\"ur Quantenoptik, Hans-Kopfermann-Str.~1, D-85748 Garching, Germany}
\affiliation{Munich Center for Quantum Science and Technology (MCQST), Schellingstr. 4, D-80799 M\"unchen}

\begin{abstract}
We develop general tools to characterise and efficiently compute  relevant observables of multimode  $N$-photon states generated in non-linear decays in one-dimensional waveguides. We then consider optical interferometry in a Mach-Zender interferometer where a  $d$-mode photonic state enters in each arm of the interferometer. We derive a simple expression for the Quantum Fisher Information in terms of the average photon number in each mode, and show that it can be saturated by number-resolved photon measurements that do not distinguish between the different $d$ modes.   
\end{abstract}
\maketitle

\section{Introduction}

Photonic states with a large and fixed number $N$ of photons play a crucial role in quantum technologies but are extremely challenging to prepare experimentally. The paradigmatic example are single-mode Fock states, $\ket{N}\propto (a^\dagger)^N\ket{0}$, where all the photons share the same spatio-temporal mode ($a^\dagger$), and which are the basis of many quantum metrology protocols~\cite{Giovannetti2011,Tth2014,Dowling2015,demkowicz-dobrzanski15}. Nowadays, the most widely used method to generate them is based on combining with post-selection heralded single photons emitted in spontaneous parametric down-conversion processes~\cite{dakna99a,ourjoumtsev06a,waks06a,zavatta08a,wang16a}. This method, however, suffers from an exponential decrease of efficiency with $N$, hindering its application for large photon numbers. Single-mode Fock states can also be  emitted naturally from entangled atomic states in ensembles with many more atoms than $N$~\cite{porras08a}. However, exciting such atomic states is highly non-trivial because of the linear energy spectrum of such systems~\cite{lemr09a,duan03a,gonzaleztudela15a,gonzaleztudela17a}.

A way of circumventing these limitations is the use non-linear systems for the generation of such photonic states. These type of systems appear in many different contexts, such as in cavities with Kerr-type non-linearities~\cite{gupta07a}, in multi-level quantum dots due to biexciton binding-energies~\cite{dousse10a,ota11aa}, or even in atomic ensembles simply because an atom can not be doubly excited or by exploiting Rydberg blockade~\cite{lukin01a,saffman10a}. These mechanisms ultimately translate into either non-harmonic energy splittings (Kerr cavity QED) or non-harmonic decay rates (saturation), which can be harnessed for multiphoton emission. An illustration of that is the proposal put forward by us in Ref.~\cite{Paulisch2019} to generate multiphoton states with quantum emitters coupled to photonic waveguides~\cite{vetsch10aaa,thompson13a,goban13a,laucht12a,faez14a,  lodahl15a,sipahigil16a,corzo16a,sorensen16a,solano17a}. There, $N$ excited emitters interact with the waveguide in the so-called mirror configuration~\cite{corzo16a,sorensen16}, such that its dynamics is described by the well-known Dicke model~\cite{dicke54a}. In that situation, the emitters experience a non-linear decay process, known as superradiant decay, which enhances the probability of emitting the photons into the waveguide as compared to other decay channels. Beyond the collective enhancement, the non-linearity has another effect: the photons released into the waveguide have an inherent \emph{multimode} structure~\cite{gonzaleztudela15a}:
\begin{align}
\ket{\phi^{(N)}_A}
=\multInt \frac{ \dd k_1 ... \dd k_N}{(2\pi)^N N!}    A_{ \{ k\}}  a^\dagger_{k_1}  \dots {a}^\dagger_{k_{N}} \ket{0},
\label{eq:multimode}
\end{align}
where $a_{k_i}^{\dagger}$ is the creation operator
of a waveguide photon of momentum $k_i$. The coefficient $A_{ \{ k \}} = A_{k_1, k_2, \cdots k_n}$ characterizes the multimodal structure of the wavepacket, and will be non-factorizable  ($A_{ \{ k \}} \neq \sqrt{N!} A_{k_1} \cdots A_{k_n}$) for photons emitted from any type of non-linear system (e.g., non-harmonic energies or non-linear decay rates). This non-trivial multimode nature of the emitted  wavepackets forces one to revisit the results derived for single-mode Fock states as they are not necessarily valid anymore. For example, the multimodal structure poses limits on the scalability as Fock state sources from spontaneous parametric down conversion processes~\cite{christ12a,tiedau19a,quesada19a}, is required to accurately predict the scattering of quantum pulses~\cite{baragiola12a,kiilerich19a}, or, as we showed in our recent manuscript~\cite{Paulisch2019}, renormalizes the results of single-mode quantum metrology protocols.

Motivated by these observations, the goal of this article is to develop general tools to deal with multimode states generated in non-linear decays, both from the point of view of its characterisation as well as applications in quantum metrology.  We start by  considering  a general $N$-dimensional emitter decaying in a non-linear fashion with a waveguide. The wavefunction  of the emitted photonic state  (given by $A_{ \{ k\}} $ in~\eqref{eq:multimode}) can be obtained through the techniques of~\cite{shi15a}, and it involves $N!$ terms due to bosonic symmetrisation. Given this highly non-trivial state, the main contributions of this article are:
\begin{enumerate}
\item  To  develop  a framework to compute relevant observables of  multimode states generated in non-linear decays in an efficient manner, with the complexity scaling polynomially with $N$. 
\item  To apply these general tools to the characterisation of Dicke superradiant photonic states where we find that most photons are contained in a few modes: a single mode contains $91\%$ of the photons, and two (three) modes already contain  $98.4\%$ ($99.6\%$) of them. 
\end{enumerate}
Having identified general properties of multimode states, and in particular superradiant Dicke states, we  then study their potential for quantum metrology \cite{Giovannetti2011,Tth2014,Dowling2015,demkowicz-dobrzanski15}. Building on our previous work~\cite{Paulisch2019}, our goal is to extend well known results in quantum optical interferometry \cite{holland93,campos03a} to the presence of a non-trivial multimode structure within the input photonic states, as in Eq.~\eqref{eq:multimode}. For that,  we consider phase estimation in a Mach-Zender interferometer, where the input state of each arm of the interferometer is a generic $d$-mode state with a fixed total photon number. Then, our main contributions are:
\begin{enumerate}
\item We show that the quantum Fisher information~\cite{Braunstein1994} (QFI) $\mathcal{Q}$ takes the particularly simple form
\begin{align}
\mathcal{Q}=  2\sum_{j=1}^d n_j (m_j+1) +m_j (n_j+1) .
\label{qfimultimodeIII}
\end{align}
where $n_i$, $m_i$ are the average photon number in the $i$th mode of the two incoming wavepackets. 
\item We show that the QFI \eqref{qfimultimodeIII} can be saturated by number-resolved measurements which \emph{cannot} distinguish between the different $d$ modes. 

\item Finally, we also  discuss the effect of  photon loss in the interferometer and in the measurement devices given the proposal of~\cite{Paulisch2019} for quantum-enhanced metrology with twin Dicke superradiant states.
\end{enumerate}

The paper is structured as follows: We start presenting the general multimode structure of the photons emitted from non-linear systems into waveguides in Section \ref{effdescriptions}, whereas the tools to characterize such photonic states are developed in Section \ref{effdescriptions}. These tools are applied to superradiant photonic states in Section \ref{superradiant_states}. In Section  \ref{quantum_metrology} we consider quantum metrology with multimode states, and finally we summarize our findings in Section \ref{conclusions}.

\section{Non-linear systems decaying in 1D waveguides}
\label{non-linear}

 \begin{figure}
	\centering
	\includegraphics[width=1\linewidth]{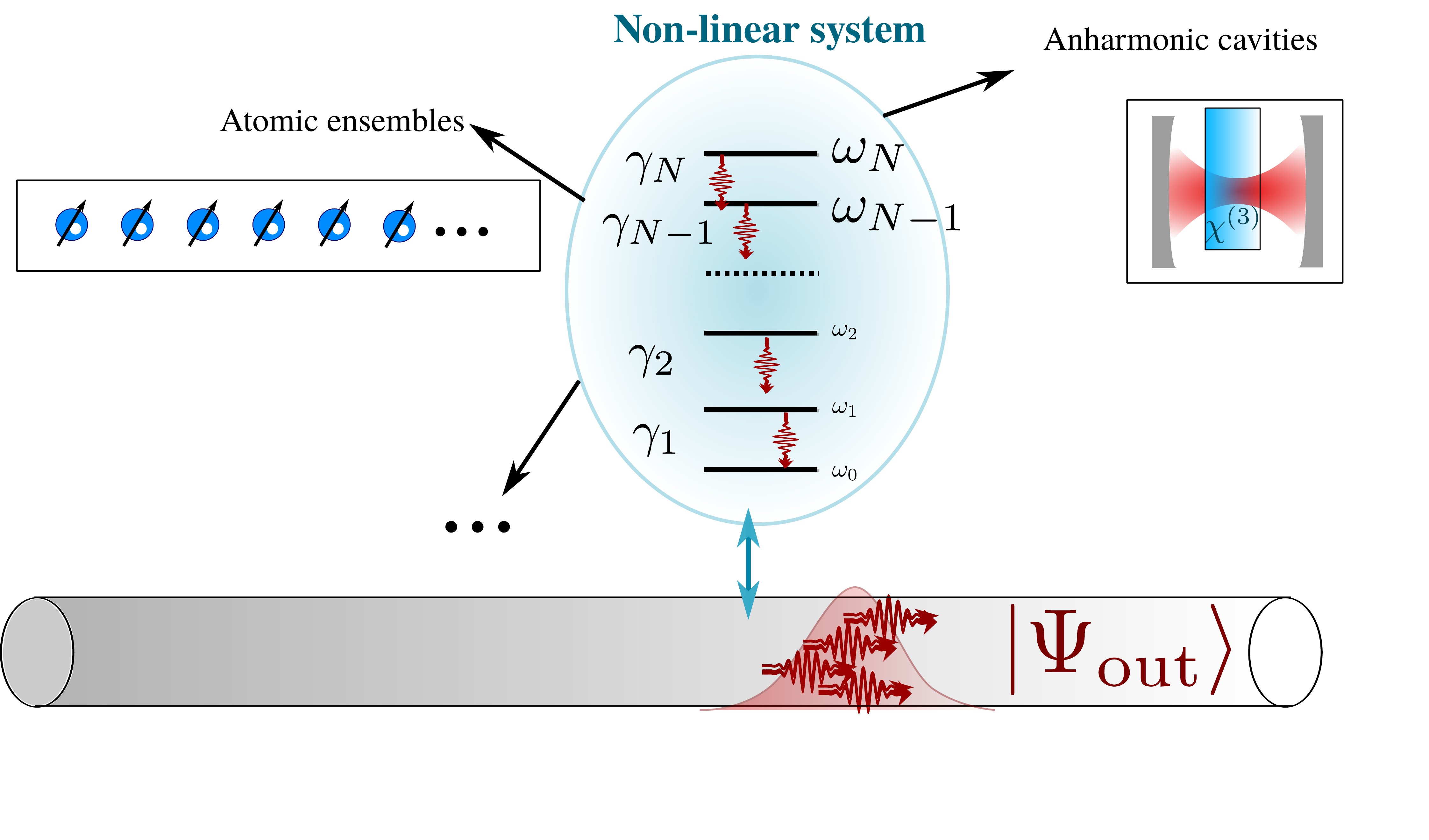}
	\caption{A the non-linear system (in blue) with $N$ levels with energy $\omega_n$, couples to a 1D waveguide. The coupling with the waveguide induce single-photon transitions $n\rightarrow n-1$ (in red arrows) at a rate $\gamma_n$. Around the non-linear system, we depict the two example of non-linear system that we consider along the manuscript, that are, equally spaced atomic ensembles and anharmonic cavities.}
	\label{setup}
\end{figure}

In the interest of generality, we consider the emission process coming from a $N$-level system ($\ket{0}, \ket{1},...,\ket{N}$) with energies $\omega_j$. Its free Hamiltonian is then given by (taking $\hbar\equiv 1$):
\begin{align}
H_S = \sum_{j=1}^{N} \omega_j \sigma_{j,j}
\end{align}
with $\sigma_{ik}= \ket{i} \bra{k}$. The system is coupled to a 1D waveguide, described by a one-dimensional and chiral photon bath with a linear dispersion (both the chirality and linearity assumptions can be relaxed obtaining similar results) $\omega_q=c q$. Taking $c\equiv 1$, its Hamiltonian read:
\begin{align}
H_B= \int  {\rm d}q \hspace{1mm} q a^{\dagger}_q a_q. 
\end{align}
Finally, the system-bath interaction Hamiltonian is assumed to be given by:
\begin{align}
H_{SB}= \sum_{j=1}^{N}\sqrt{\gamma_j} \int  \frac{d q}{2\pi} \left(a^{\dagger}_q \sigma_{j-1,j}+a_q \sigma_{j,j-1} \right).
\end{align}
where $\gamma_j$ denotes the decay rate of the transition from the $j$-th to the $(j-1)$-th level. The global Hamiltonian describing the emission process is then given by the sum of the three terms: $H=H_S+H_B+H_{SB}$. The whole physical set-up is illustrated in Figure \ref{setup}.

We consider that initially the waveguide B is in the ground state, whereas the system S initially contains $N$ excitations. When the excitations decay into the waveguide, the photonic state is described by the wavefunction (naturally extending the considerations of \cite{gonzaleztudela15a}):
\begin{align}
\ket{\phi^{(N)}}=\frac{1}{N!}\multInt_0^{\infty} A_{t_1 ... t_N }  \prod_{j=1}^N {\rm d} t_j \hspace{1mm}   a^{\dagger}_{t_j} \ket{0}
\label{genstate}
\end{align}
with
\begin{align}
A_{t_1 ... t_N}= \mathcal{T}\left(\bra{\varphi_0}\mathcal{O}_{t_1}...\mathcal{O}_{t_N}\ket{\varphi_N}\right)
\label{A}
\end{align}
where $\mathcal{T}$ stands for time-ordering, and  $\mathcal{O}_{t}$ satisfies
$\bra{\varphi_{j-1}}\mathcal{O}_{t}=\sqrt{\gamma_{j}}e^{\left[i(\omega_{j-1}-\omega_j)+\frac{1}{2}(\gamma_{j-1}-\gamma_{j})\right]t}\bra{\varphi_{j}}.$
Furthermore,  the bosonic creation and annihilation operators $a_s,a^{\dagger}_s$ satisfy the standard commutation relation
\begin{align}
[a_s,a_t^{\dagger}]=\delta(s-t).
\label{comm}
\end{align}

The use of this general light-matter Hamiltonian $H$ allows to capture the physics of very different models, such as:
\begin{itemize}
	\item \emph{Saturated atomic ensembles in the atomic mirror configuration}. As explained in Refs.~\cite{gonzaleztudela15a,Paulisch2019}, within the Markov approximation the coupling of the ensemble with the waveguide can be described by a single collective dipole operator. This can be effectively described as $N$-level system with equally spaced energy levels, $\omega_n=n\omega_0$, but non-linear decay rates: $\gamma_n=\Gamma_{\mathrm{1d}} n(N-n+1)$. In the text, we shall call these emitted states \emph{superradiant states}, or \emph{superradiant photonic states}. 
	
	\item \emph{Anharmonic cavities}. In the case of a non-linear resonator, with a Kerr non-linearity of the form $H_B=\omega_a a^\dagger a+U a^\dagger a (a^\dagger a-1)$, but coupled to the waveguide in a linear fashion such that its reduced dynamics is given by the standard Lindblad form $\Gamma_{\mathrm{1d}}(2a\rho a^\dagger-a^\dagger a\rho-\rho a^\dagger a)/2$, the energy levels are now non-linear: $\omega_n =n\omega_a+n(n-1) U$, while the decay rates are harmonic $\gamma_n=n\Gamma_{\mathrm{1d}}$.
	
\end{itemize}
 
Along this manuscript, we will focus on the characterization of the first ones, as they will be the most relevant for quantum metrology. However, all the formalism developed is valid for any combination of $\{\omega_n,\gamma_n\}$.

\section{Towards effective characterisations of multimode states}
\label{effdescriptions}

We start this section by developing techniques that enable us to compute relevant observables of $\ket{\phi^{(N)}}$ efficiently for large $N$. This is motivated by noting that the analytical form  \eqref{genstate} contains $N!$ terms due to the  time ordering in \eqref{A}, making such an expression  challenging to handle beyond low $N$. 

We first consider a single emitted photon   (with $\gamma \equiv \gamma_1-\gamma_0$ and $\omega\equiv \omega_1 -\omega_0$):
\begin{align}
\ket{\phi^{(1)}}=\int_0^{\infty} {\rm d} t B_t^{(\gamma,\omega)}  a^{\dagger}_{t} \ket{0}\equiv b_{\gamma,\omega}^{\dagger} \ket{0}
\end{align}
where we have defined $B_t^{(\gamma,\omega)} \equiv \sqrt{\gamma} e^{-t\left(i  \omega +\frac{\gamma}{2}\right)}$ and  $b_{\gamma,\omega}^{\dagger}\equiv \int_0^{\infty} {\rm d} t B_t^{(\gamma,\omega)}  a_{t}^{\dagger}$.
Using \eqref{comm}, the commutation relation
\begin{align}
[b_{\gamma,\omega},b^{\dagger}_{\gamma',\omega'} ]=\frac{2\sqrt{\gamma  \gamma'}}{\gamma+  \gamma'+2i(\omega'-\omega)}
\label{commxymain}
\end{align}
follows. Given such single-mode operators $b_{\gamma,\omega}$'s, we have developed techniques to compute efficiently observables of the form 
\begin{align}
\bra{\phi^{(N)}}b_{x_1,y_1} ... b_{x_n,y_n} b_{\tilde{x}_1,\tilde{y}_1}^{\dagger} ... b_{\tilde{x}_n,\tilde{y}_n}^{\dagger}\ket{\phi^{(N)}}.
\label{generalexpressionfmt}
\end{align}
The computational techniques for dealing with \eqref{generalexpressionfmt} are rather involved and are developed in detail in the Appendix  \ref{RecurrenceRelations}. Here, we instead explain the main ideas and implications:
\begin{enumerate}
\item The computation of \eqref{generalexpressionfmt} is developed by expressing it as a recurrence relation, which is then transformed into a matrix multiplication. 
\item The solution is exact as the integrals in \eqref{genstate}  are carried out analytically. Yet, in practice it is convenient to perform the matrix multiplication numerically in order to access large $N$.
\item The size of the involved matrices is at most $(4^{n} (N-n+1))^2$, although usually this can be reduced if there are some symmetries in the calculation (i.e. if some of the $x_i$'s and $y_i$'s in \eqref{generalexpressionfmt} are the same).   In practice, this means that for  $n=1$ (corresponding to average photon number), one can easily reach up to $N\approx 1000$, whereas for $n=2$ (corresponding to the variance) one can reach $N\approx 100$. This should be contrasted to the naive calculation of \eqref{generalexpressionfmt} from \eqref{genstate} which involves $(N!)^2$ integrals.
\item Since the  $b_{\gamma,\omega}^{\dagger}$'s form  an (overcomplete) basis of the Fock space spanned by $a^{\dagger}_t$ $\forall t$, one can in principle compute arbitrary observables through this approach. 
\end{enumerate}

\section{Characterisation of superradiant photonic states}
\label{superradiant_states}

We now apply the  machinery developed in the previous section to the characterisation of superradiant photonic states emitted  when $N$ excited atoms are placed next to the waveguide in the atomic mirror configuration, as previously proposed by us in~\cite{Paulisch2019}. This corresponds to taking $\ket{\phi^{(N)}}$ in \eqref{generalexpressionfmt} with $\gamma_n=\Gamma_{\mathrm{1d}} n(N-n+1)$ and $\omega_n=n\omega_0$.

Before proceeding to its characterisation, let us mention that there are two main features that make this proposal particularly appealing \cite{Paulisch2019}:
\begin{itemize}
\item In the absence of photon loss (i.e., when the atom-waveguide set-up is perfectly isolated) and assuming perfect control on the system's Hamiltonian, the protocol is \emph{deterministic} and \emph{scalable}. That is, simply by placing more atoms next to the waveguide, one can generate a larger $N$-photon state.
\item In the presence of photon loss in free space, the probability of success scales as $\approx 1-\ln( N) (\Gamma^*/\Gamma_\oned)$, where $\Gamma_\oned$ is the decay rate into the waveguide and $\Gamma^*$ into free space, for $\Gamma^*/\Gamma_\oned\ll 1$. This slow decrease with $\ln N$ arises due to the enhanced collective decay and  should be contrasted with the standard  success probability $\approx 1- N (\Gamma^*/\Gamma_\oned)$  obtained for $N$ independent decay processes (see Ref. \cite{Paulisch2019} for more details).  
\end{itemize}
Hence, Dicke superradiant decays  provide a natural framework to generate $N$-photonic states in a deterministic, scalable and robust-to-photon-loss manner. Furthermore, current experimental nanophotonic platforms have already achieved ratios $\Gamma_\oned/\Gamma^*\approx 60$ with  $\Gamma_\oned\sim 1$ GHz~\cite{lodahl15a}.

\subsection{ Average photon number }
\label{Sec:AvPhotNum}
\begin{figure}
	\centering
	\includegraphics[width=1\linewidth]{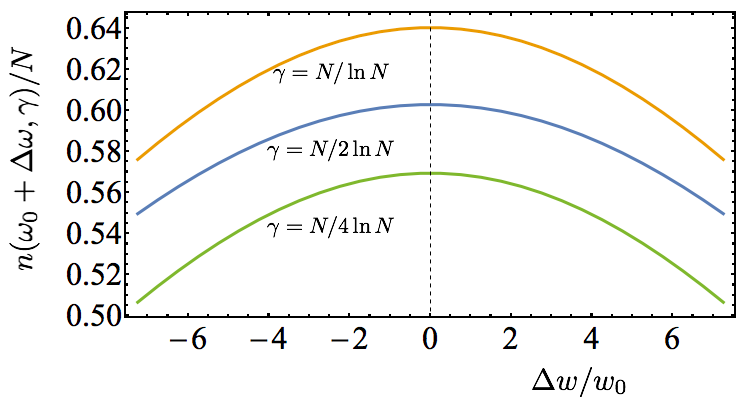}
	\caption{Ratio of photons of a superradiant   $N$-photon state $\ket{\phi^{(N)}}$ in a mode with frequency $\omega_0+\Delta \omega$ and inverse decay rate $\gamma$.}
	\label{avPhotonII}
\end{figure}

  We first consider the photon number 
\begin{align}
n(\gamma,\omega)\equiv \bra{\phi^{(N)}}b_{\gamma,\omega}^{\dagger}  b_{\gamma,\omega} \ket{\phi^{(N)}}.
\end{align} 
We start by fixing the decay rate $\gamma$ and varying $\omega$ around $\omega_0$, corresponding to the frequency of the two-level emitters.  Figure \ref{avPhotonII} shows how the average photon number  is centred at $\omega_0$, as one would have expected physically. 

\begin{figure}
	\centering
	\includegraphics[width=1\linewidth]{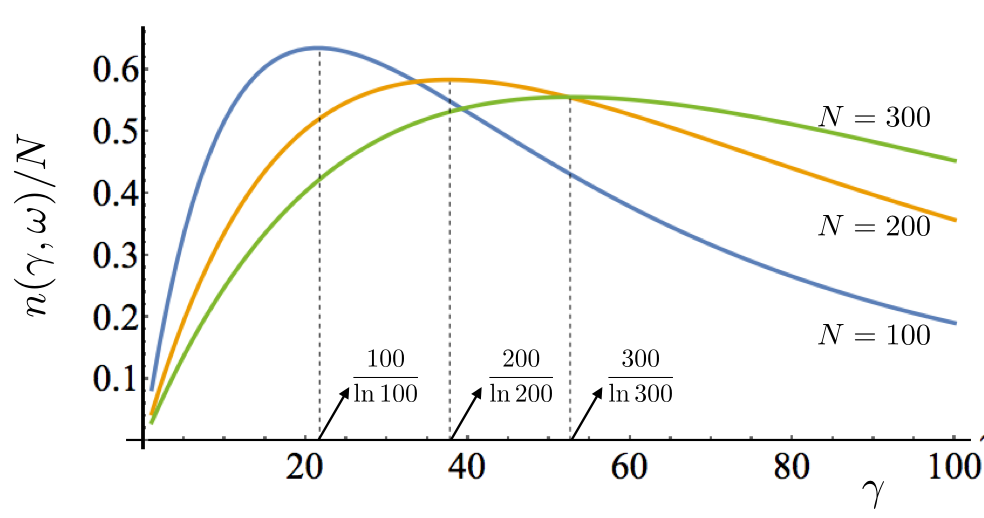}
	\caption{Ratio of photons of a superradiant   $N$-photon state $\ket{\phi^{(N)}}$ in a mode with frequency $\omega_0$ and varying inverse decay rate $\gamma$. Note that the maximal number of photons is found for $\gamma \approx N/\ln N$. }
	\label{avPhoton}
\end{figure}

Next we consider  $n(\gamma,\omega)$ for a fixed $\omega=\omega_0$ and varying $\gamma$. The results are shown in  Fig. \ref{avPhoton}, where we compute $n(\gamma,\omega_0)$ for $N=100, 200, 300$. Interestingly, note that  the maximum of $n(\gamma,\omega)$ appears at $x=N/\ln N$.
 To understand this,  note that the time scale $\tau$ of decay of the   photons is proportional to $\gamma^{-1}$. In particular, for the superradiant decay, the $j$th collective exitation decays with a time scale $ \gamma_j^{-1}$, with $\gamma_j=\Gamma j(N-j+1)$, and the average decay time is given by
\begin{align}
\langle \tau \rangle  \propto \sum_{j=1}^{N} \gamma_j^{-1}=\Gamma^{-1}\sum_{j=1}^{N} \frac{1}{j(N-j+1)}\approx \Gamma^{-1}\frac{N}{2 \ln N}.
\end{align}
 Hence, we realise from $\eqref{genstate}$ that the choice $\gamma=N/\ln N$ corresponds to single-mode photons decaying with the average decay time of the superradiant photons. This provides an heuristic explanation for the optimal choice of $\gamma$  that maximises  $n(\gamma,\omega_0)$, i.e., the number of photons in the mode  $ b_{\gamma,\omega} $.
 
\subsection{Most relevant modes}
\label{MostRelevantModes}

The results of Figures \ref{avPhoton} and \ref{avPhotonII} suggest that a rather large proportions of photons (between $50\%$ and  $60\%$ for $N=100,200,300$) can be contained in a small set of modes centred around the frequency $\omega$ and with inverse decay rate $\gamma=N/\ln N$. This motivates us to consider a set of $D$ modes with frequency $\omega_0$ and varying $\Gamma=j N/\ln N$, with $j=1,...,D$; that is, $\{b_{jx,\omega_0}\}_{j=1}^D$ with $x=N/\ln N$. 
Note that these modes are not orthogonal due to \eqref{commxymain}.  A set of orthogonal modes can be constructed by solving the generalised eigenvalue equation,
\begin{align}
T \vec{v}^{(k)} = \lambda_k R \vec{v}^{(k)}
\label{geneigeq}
\end{align}   
where $T$ and $R$ are matrices of size $D^2$ whose elements are given by
\begin{align}
&T_{kl}= \bra{\phi^{(N)}} b_{kx,\omega_0}^{\dagger} b_{lx,\omega_0} \ket{\phi^{(N)}},
\nonumber\\
&R_{kl}= [b_{kx,\omega_0},b^{\dagger}_{lx,\omega_0}]=\frac{2\sqrt{kl}}{k+l}.
\end{align}
This leads to a set of $D$ bosonic modes 
\begin{align}
\label{defd_k}
c_k=\sum_{j=1}^D v^{(k)}_j b_{jx,\omega_0} \hspace{5mm} {\rm with } \hspace{5mm} x=\frac{\ln N}{N},
\end{align}
The usual commutation relations
\begin{align}
[c_i,c^{\dagger}_j]=\delta_{ij},
\label{commds}
\end{align}
are guaranteed by \eqref{geneigeq} (plus appropriate normalisation). 

 \begin{figure}
	\centering
	\includegraphics[width=1\linewidth]{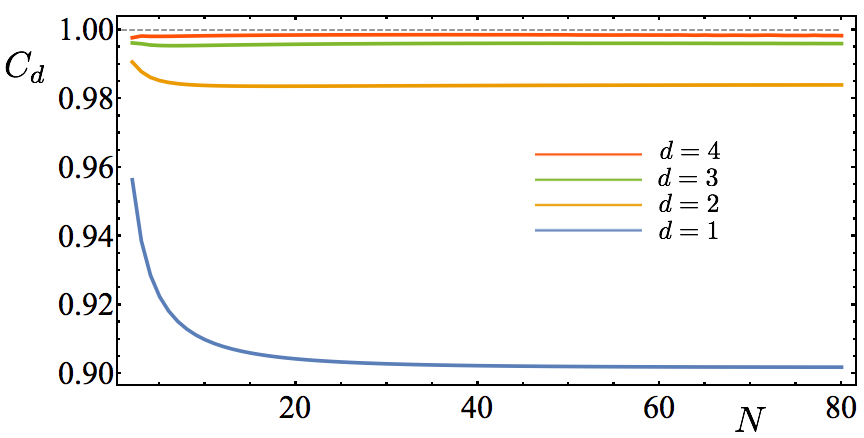}
	\caption{Proportion  of photons in $d$ modes as quantified by \eqref{C_d} for $D=10$ as a function of the number of photons $N$ in the superradiant state.}
	\label{avPhotonIII}
\end{figure}

Let us now define the number operators (for a given $D$),
\begin{align}
&n_d\equiv \sum_{j=1}^d c_j^{\dagger}c_j,
\label{n_d}
\end{align}
and the corresponding ratio of photons,
\begin{align}
C_{d}=\frac{ \bra{\phi^{(N)}} n_d \ket{\phi^{(N)}}}{N}.
\label{C_d}
\end{align}
In Figure \ref{avPhotonIII},   we show $C_d$ 
as a function of $N$ for $D=10$ (i.e. the modes $c_k$'s are a linear combination of 10 modes $b_{kx,\omega_0}$'s). Note  that $C_d$ quickly saturates with $N$. 
The values of $C_d$ for different $D$'s are shown in the following Table, which is evaluated at $N=100$:
\begin{center}
\begin{tabular}{ ||c||c|c|c|c|c|c|c|c|c|c|| } 
 \hline
  \hline
  & $C_1$ & $C_2$& $C_3$ & $C_4$  & $C_5$ & $C_6$& $C_7$ & $C_8$ \\ 
 \hline
 \hline
 D=2 &0.653 & 0.688 &  & &  &  & &   \\ 
 \hline
 D=4 &0.852 & 0.918 & 0.921  & 0.921  & &  &  &\\ 
 \hline
 D=6 &0.894 & 0.973 & 0.980  & 0.981 & 0.981 & 0.981 &   &   \\ 
 \hline
 D=8 &0.901 & 0.983 & 0.993  & 0.995 & 0.995 & 0.995 & 0.995 & 0.995 \\ 
 \hline
 D=10 &0.902 & 0.984 & 0.996  & 0.998 &0.999 & 0.999 & 0.999  & 0.999 \\ 
  \hline
   \hline
\end{tabular}
\label{Table}
\end{center}
These numbers can be slightly increased by considering larger $d$'s or $D$'s.  It is remarkable that with only 2 (3) modes we can cover $98.4\%$ ($99.6\%$) of the photons, and that more than $90\%$ photons live in a single mode. Hence, although the superradiant state   $\ket{\phi^{(N)}}$ may naively appear as a highly multimode state, it can be described by means of only a few modes. At the same time, it is worth noticing that although we reached this result by a rather heuristic method, these descriptions are almost optimal: since the small set of considered modes already contains around $99.9\% $ of the photons, it is not possible that by considering a much larger (possibly infinite) set of modes the effective descriptions can considerably change. In other words, while a few modes seem to suffice to describe $\ket{\phi^{(N)}}$ with high accuracy, it is  not possible  to describe it by a single one.

\subsection{ Variance}
\begin{figure}
	\centering
	\includegraphics[width=1\linewidth]{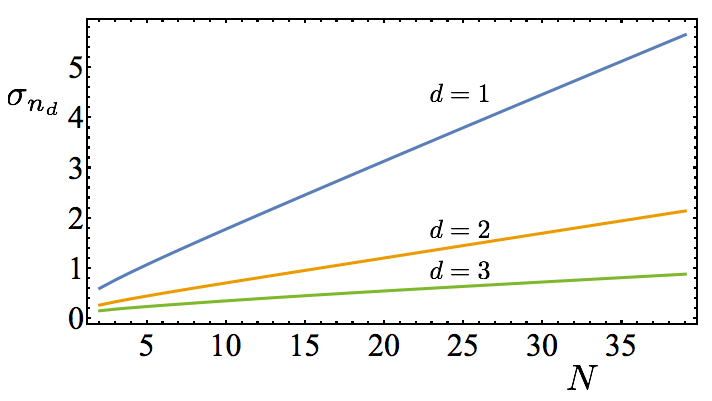}
	\caption{Variance of the number operators  $n_1$, $n_2$, and $n_3$ defined in \eqref{n_d} for $D=8$.}
	\label{var1}
\end{figure}

 Let us further characterise the fact that most photons are contained in a few modes by studying  the fluctuations of  the number operators  $n_k$'s given in \eqref{n_d}. In particular  consider the variance 
\begin{align}
\sigma_{n_d}=\sqrt{ \langle n_d^2 \rangle -\left(\langle  n_d \rangle \right)^2},
\end{align} 
with $\langle ... \rangle  \equiv \bra{\phi^{(N)}} ... \ket{\phi^{(N)}}$.
To compute such expressions, first we expand them using the commutation relations \eqref{commds} as
\begin{align}
\langle n_1^2 \rangle =&\langle c_1^2 (c^{\dagger}_1)^2  \rangle -3 \langle n_1 \rangle -2 -\left(\langle n_1 \rangle \right)^2  
\nonumber\\
\langle n_2^2 \rangle = &\langle  c_1^2 (c^{\dagger}_1)^2  \rangle+\langle c_2^2 (c^{\dagger}_2)^2 \rangle + 2 \langle c_1 c_2 c_1^{\dagger} c_2^{\dagger} \rangle 
\nonumber\\
 &-5 \langle n_1 \rangle-5 \langle n_2 \rangle -6  
\end{align}
and similarly for higher $\langle n_j^2 \rangle$. Each term can be evaluated by using \eqref{defd_k} and then computing  $\langle b_i b_j b^{\dagger}_k b^{\dagger}_l \rangle $  (or $\langle b_i b_j \rangle $) by the appropriate recurrence relation derived in Appendix \ref{RecurrenceRelations}. For numerical purposes, it is convenient to first compute  $\langle b_i b_j b^{\dagger}_k b^{\dagger}_l \rangle $ $\forall i,j,k,l$ where $i=1,...,D$ and save the results. In the numerical simulations of this section, we take $D=8$, for which we already cover $99\%$ of the photons as shown in Table \ref{Table}.   The results are shown in Figure \ref{var1}  for $d=1,2,3$ and $D=8$. As expected, $\sigma_{n_d}$ decreases as $d$ increases, and for $d=3$ and $D=8$ the variance is rather small: less than one photon for $N=40$.  This confirms the previous results that a few modes can cover most of the photons contained in the superradiant state.  Furthermore, we also  observe that $\sigma_{n_d}$ grows linearly with $N$, which is compatible with the proportion of photons staying constant with $N$ as shown in Figure \ref{avPhotonII}. 

\subsection{Effective descriptions}
\label{EffectiveDescriptions}

The main insight of the previous sections is that 2 or 3 modes suffice to describe most of the photons of $\ket{\phi^{(N)}}$. We can use this insight to build effective descriptions of $\ket{\phi^{(N)}}$ within these subspaces. This can be achieved by computing the overlaps
\begin{align}
\label{overlapdsmain}
\langle 0 | c_1^{k_1}c_2^{k_2}... c_d^{k_d}  \ket{\phi^{(N)}},
\end{align}
for all $\{c_j\}_{j=1}^{d}$ such that $\sum_{d} k_d=N $, and where $d$ is the number of considered modes. For that, we first express \eqref{overlapdsmain} as a combination of $\langle 0 | b_{x,\omega_0}^{k_1}b_{2x,\omega_0}^{k_2}... b_{Dx,\omega_0}^{k_D}  \ket{\phi^{(N)}}$ through \eqref{defd_k}. Then,  in Appendix \ref{RecurrenceRelations}, we develop a recurrence relation  method to compute $\langle 0 | b_{x,\omega_0}^{k_1}b_{2x,\omega_0}^{k_2}... b_{Dx,\omega_0}^{k_D}  \ket{\phi^{(N)}}$
which requires the multiplication of $\mathcal{O}(N)$ matrices of dimension ${\rm dim}$, which satisfies  $N \lesssim {\rm dim} \lesssim \left( N/D \right)^D$. 
In order to compute \eqref{overlapdsmain}, it is convenient to first  compute all possible $\langle 0 | b_{x,\omega_0}^{k_1}b_{2x,\omega_0}^{k_2}... b_{Dx,\omega_0}^{k_D}  \ket{\phi^{(N)}}$ for all $\{k_j\}_{j=1}^N$ such that $\sum_{j} k_j=N $  and save such coefficients. There are $\mathcal{O}(N^D)$ such coefficients, which provides the necessary space to carry out the computation. Overall, the whole computation is challenging,  both in terms of the number of operations and the complexity of each one, but becomes feasible for small $N$ and $D\leq 10$, which is enough to capture most of the photons (see Table \ref{Table}).

\begin{figure}
	\centering
	\includegraphics[width=1\linewidth]{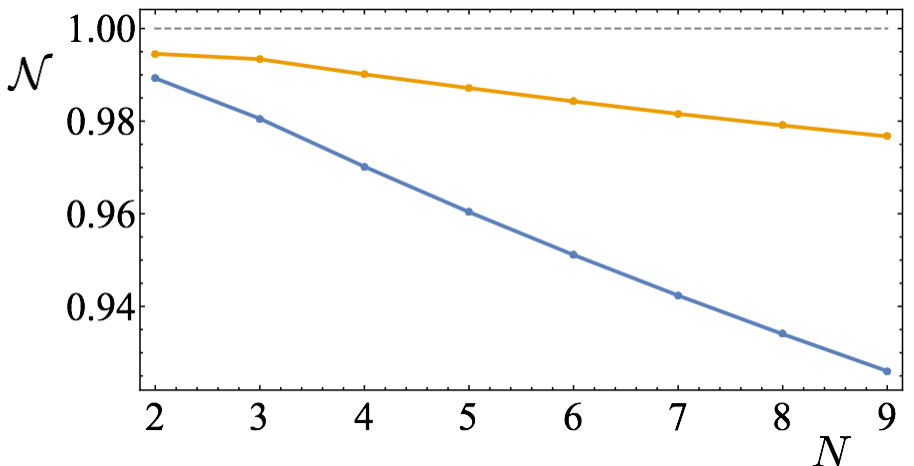}
	\caption{Overlap for the two-mode (blue) and three-mode (orange) approximations for $D=8$.}
	\label{norm}
\end{figure}

We  consider the following unnormalised two and three mode state:
\begin{align}
&|\psi^{(N)}_2\rangle = \sum_{j=0}^N \alpha_j \ket{j,N-j}  
\nonumber\\
&|\psi^{(N)}_3\rangle = \sum_{j=0}^N \sum_{k=0}^{N-j} \alpha_{j,k} \ket{j,k,N-j-k}  
\label{psi}
\end{align}
with $\ket{j,N-j} = (c_1^{\dagger})^{N-j} (c_2^{\dagger})^{j} \ket{0}/\sqrt{(N-j)!j!}$, $\hspace{1mm}\ket{j,k,N-j-k} = (c_1^{\dagger})^{j} (c_2^{\dagger})^{k}  (c_3^{\dagger})^{N-j-k}\ket{0}/\sqrt{(N-j-k)!k!j!}$, $\alpha_j=\braket{\phi^{(N)}}{j,N-j} $ and $\alpha_{j,k}=\braket{\phi^{(N)}}{j,k,N-j-k} $. 
The normalisation (for two modes)
\begin{align}
\mathcal{N}=\sum_{j=0}^N |\alpha_j|^2, 
\end{align}
or $\mathcal{N}=\sum_{j,k} |\alpha_{j,k}|^2$ for three,
indicates the overlap between $|\psi^{(N)}_{2,3}\rangle$ and $|\phi^{(N)}\rangle$, i.e., how good the approximation in the two or three mode subspace  is. In Figure \ref{norm}, we show how $\mathcal{N}$ is close to 1 for low $N\leq 10$, especially for $d=3$ where it stays above 0.97. This implies that $\ket{\phi^{(N)}}$ can indeed be well described by two or three mode states as in \eqref{psi}. For $d=2$, the coefficients are given in Figure \ref{coefII} and the coefficients for  $d=3$ are provided in Appendix \ref{RecurrenceRelations}. These approximate states of the form \eqref{psi} become  handy in calculations for quantum information tasks, as we will later illustrate for quantum metrology, and importantly our techniques enable us to quantify how close are our approximate states to the real $|\phi^{(N)}\rangle$. As a final remark, we note that the linear decrease of $\mathcal{N}$ with $N$ is also compatible with our previous results where we observed that the proportion of photons in a given mode stays constant. 

 \begin{figure}
	\centering
	\includegraphics[width=1\linewidth]{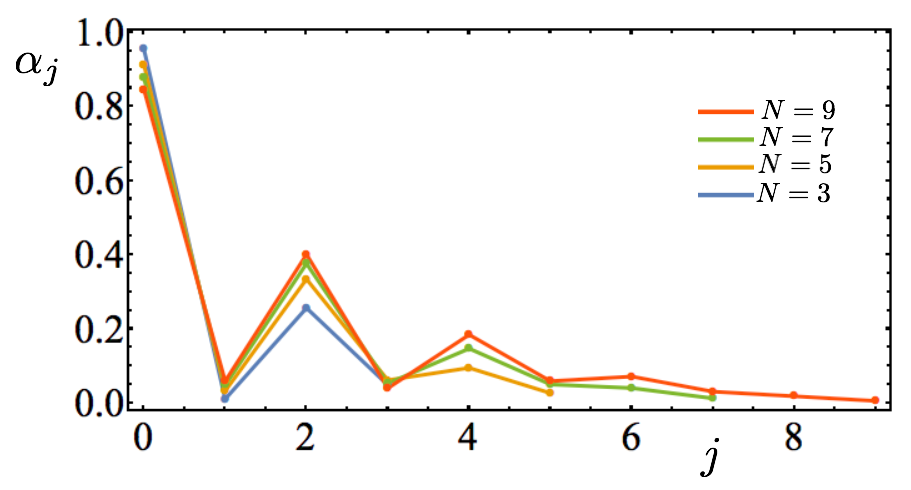}
	\caption{Coefficients of the two-mode state for $N=3,5,7,9$ and $D=8$.}
	\label{coefII}
\end{figure}

\section{Applications in quantum metrology}
\label{quantum_metrology}

\begin{figure}
	\centering
	\includegraphics[width=1\linewidth]{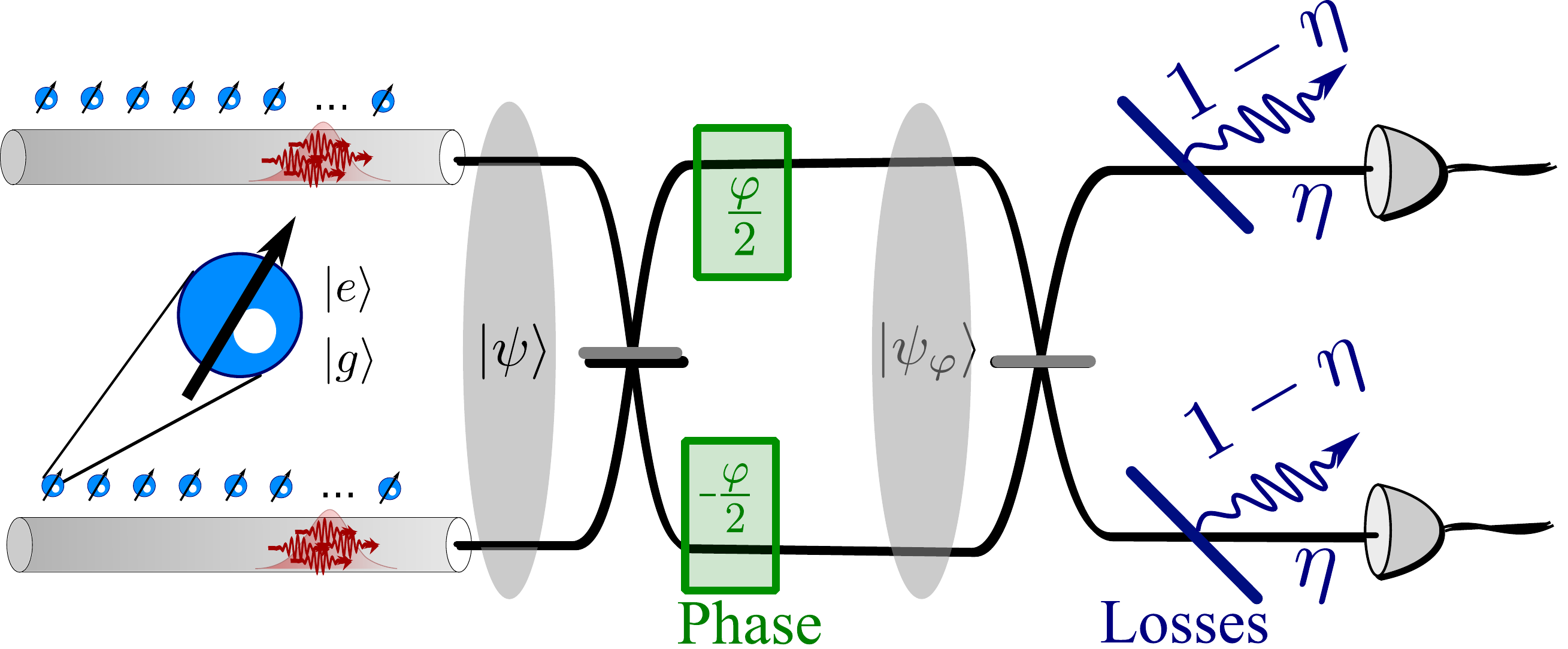}
	\caption{A multimode state (generated e.g. through a superradiant decay) enters  each arm of a MZ interferometer. Each arm of the interferometer is described by the set of modes: $\{a_k \}$, $\{ b_k\}$.   }
	\label{MZscheme}
\end{figure}

We now apply the different insights and techniques developed in the last section to  quantum optical interferometry, in particular by considering phase estimation in a standard Mach-Zender (MZ) interferometer. We consider that each arm of the MZ inteferometer  is described by a set of modes $\{a_k\}_{k=1}^d$ and $\{b_k\}_{k=1}^d$, respectively (see Figure \ref{MZscheme}). When $d=1$, we recover the standard two-mode optical interferometry \cite{demkowicz-dobrzanski15}, and our goal is precisely to extend well-known results to the multimode regime where $d>1$.  We start this section by introducing some basic concepts of quantum metrology, as well as describing the set-up we consider here in detail. 

We consider the estimation of a parameter $\varphi$ by measuring $N$ photons which encode information about it. The (possibly entangled) $N$-photon state is described by $\ket{\psi_{\varphi}}$.  Let us assume that we apply a measurement $M$ on $\ket{\psi_{\varphi}}$, the statistics being described by a probability distribution $P(s_j|\varphi)$ where $\{s_j\}$ are the possible outcomes of the measurement given the value $\varphi$ of the unknown parameter to be estimated.   If this  process  is repeated $\nu$ times,  in the limit $\nu \gg 1$   the Cramer-Rao bound~ guarantees that the mean-squared error $\Delta^2 \tilde{\varphi}$ of any unbiased and consistent estimator $\tilde{\varphi}$ of $\varphi$ is lower bounded by~\cite{helstrom76,holevo82}:
\begin{align}
\Delta^2 \tilde{\varphi} \geq \frac{1}{\nu \mathcal{C}}
\end{align}
where $\mathcal{C}$ is the classical Fisher Information (CFI),
\begin{align}
\mathcal{C}= \sum_{j} \frac{1}{P(s_j|\varphi)} \left( \frac{\partial P(s_j|\varphi)}{\partial \varphi} \right)^2
\label{cfi}.
\end{align}
The CFI quantifies the best resolution of a particular estimation scheme as defined through $\ket{\psi_{\varphi}}$ and a measurement $M$. In their seminal work, Braunstein and Caves optimised $\mathcal{C}$ over all possible quantum measurements $M$ \cite{Braunstein1994}. The resulting quantity is the Quantum Fisher Information  $\mathcal{Q}$ (QFI)  , which satisfies,
\begin{align}
\Delta^2 \tilde{\varphi} \geq \frac{1}{\nu \mathcal{C}} \geq \frac{1}{\nu \mathcal{Q}}
\end{align}
and, for pure states,  is given by
\begin{align}
\mathcal{Q}=4\left(\braket{\dot{\psi}_{\varphi}}{\dot{\psi}_{\varphi}}-|\braket{\dot{\psi}_{\varphi}}{\psi_{\varphi}}|^2 \right)
\label{qfi}
\end{align}
with $\dot{\psi}_{\varphi}=\partial_{\varphi}\psi_{\varphi}$. The QFI quantifies the potential of a particular state $\ket{\psi_{\varphi}}$ for quantum metrology \cite{Braunstein1994}.

Let us now discuss how we encode $\varphi$ in the standard  Mach-Zender (MZ) interferometer, but extended to the multimode regime. We consider as an initial state $\ket{\psi}= \ket{\phi_A,\phi_B} $, 
where 
\begin{align}
&\ket{\phi_A}= \sum_{\{\alpha_i \}_{i=1}^d \in \mathcal{C}(n)} C_{\{\alpha\}}^{(A)}\ket{\{\alpha\}}_a
\nonumber\\ 
&\ket{\phi_B}= \sum_{\{\alpha_i \}_{i=1}^d \in \mathcal{C}(m)} C_{\{\alpha\}}^{(B)}\ket{\{\alpha\}}_b  
\end{align}
with 
\begin{align}
\label{defabket}
  \ket{\{\alpha\}}_a= \frac{(a_1^{\dagger})^{\alpha_1}}{\sqrt{\alpha_1!}}...\frac{(a_d^{\dagger})^{\alpha_d}}{\sqrt{\alpha_d!}}\ket{0}
  \nonumber\\
  \ket{\{\alpha\}}_b= \frac{(b_1^{\dagger})^{\alpha_1}}{\sqrt{\alpha_1!}}...\frac{(b_d^{\dagger})^{\alpha_d}}{\sqrt{\alpha_d!}}\ket{0},
\end{align}
and where $\{\alpha_i \}\in\mathcal{C}(n)$ iff $\sum_i \alpha_i=n$, and with $n+m=N$. That is, we consider that two independent (but generic) states  of $m$ and $n$  photons, each of them described by $d$ modes, enter the two arms of the interferometer. For $d=1$, the states are single-mode states and hence Fock states. We also assume   bosonic commutation relations,
\begin{align}
[a_i,a^{\dagger}_j]=\delta_{ij},
\hspace{6mm}
[b_i,b^{\dagger}_j]=\delta_{ij},
\label{crelations}
\end{align}
which can always be satisfied by a proper choice of the initial modes through the generalised eigenvalue equation, as in  \eqref{geneigeq}.
The initial state $\ket{\psi}= \ket{\phi_A,\phi_B} $ goes through a balanced beam splitter $U_{BS} $, gains a relative phase $\varphi$ when traveling through the two arms of the interferometer, and finally enters another beam splitter; the final state then reads
\begin{align}
\ket{\psi_\varphi}=U_{BS}^{\dagger}e^{-i\varphi H} U_{BS} \ket{\phi_A,\phi_B} 
\end{align}
with  $H=\frac{1}{2}\left(\sum_{i=1}^d a^{\dagger}_i a_i-b^{\dagger}_i b_i\right)$, and where the transformation $U_{BS}^{\dagger}e^{-i\varphi H} U_{BS}$ can be described by:
\begin{align}
\begin{pmatrix} 
\tilde{a}_j  \\
\tilde{b}_j   
\end{pmatrix}
=
\frac{1}{2}\begin{pmatrix} 
1 & i \\
i & 1  
\end{pmatrix}\begin{pmatrix} 
e^{i\varphi/2} & 0 \\
0 & e^{-i\varphi/2}  
\end{pmatrix}\begin{pmatrix} 
1 & -i \\
-i & 1  
\end{pmatrix}
\begin{pmatrix} 
a_j  \\
b_j   
\end{pmatrix}
\end{align}
where $\tilde{a}_j , \tilde{b}_j $ are the output modes. 
 Using this transformation we easily  obtain,
\begin{align}
\tilde{H}\equiv U_{BS}^{\dagger} H U_{BS}=\frac{i}{2}\sum_{i=1}^d \left( b_i^{\dagger}a_i-a_i^{\dagger}b_i\right).
\label{movedH}
\end{align}
On the other hand,  $\ket{\psi_{\varphi}}$ can be explicitly written as:
\begin{align}
&\ket{\psi_{\varphi}}= \sum_{\{\alpha_i \} \in \mathcal{C}(n), \hspace{2mm}  \{\beta_i \} \in \mathcal{C}(m)} \frac{ C_{\{\beta\}}^{(B)} C_{\{\alpha\}}^{(A)}}{\sqrt{\prod_i \alpha_i! \beta_i!}}
\nonumber\\ 
&\left(a_1^{\dagger}(\varphi) \right)^{\alpha_1}...\left(a_D^{\dagger}(\varphi) \right)^{\alpha_D} \left(b_1^{\dagger}(\varphi) \right)^{\beta_1}...\left(b_D^{\dagger}(\varphi) \right)^{\beta_D}  \ket{0}
\end{align}
with 
\begin{align}
&a_j^{\dagger}(\varphi) =\cos(\varphi/2) a_j^{\dagger}+ \sin(\varphi/2) b_j^{\dagger}
\nonumber\\
&b_j^{\dagger}(\varphi) =-\sin(\varphi/2) a_j^{\dagger}+  \cos(\varphi/2) b_j^{\dagger}.
\end{align}
Finally, it will be useful to note that
\begin{align}
&\dot{a}^{\dagger}_j (\varphi)= \frac{1}{2} b^{\dagger}_j(\varphi),
\hspace{6mm}
&\dot{b}^{\dagger}_j (\varphi)= \frac{1}{2} a^{\dagger}_j(\varphi).
\label{derab}
\end{align}

\subsection{Quantum Fisher Information for pure multimode states}

When $\ket{\psi}=e^{-i\varphi H}$ in \eqref{qfi},  then we have the convenient expression for the QFI:
$\mathcal{Q}=4\left(\bra{\psi_{\varphi}}H^2 \ket{\psi_{\varphi}}-(\bra{\psi_{\varphi}}H\ket{\psi_{\varphi}})^2 \right)$.
Using \eqref{crelations} and \eqref{movedH}, we obtain
\begin{align}
\mathcal{Q}=&\bra{\phi_A,\phi_B}\tilde{H}^2 \ket{\phi_A,\phi_B}-\left(\bra{\phi_A,\phi_B}\tilde{H}\ket{\phi_A,\phi_B}\right)^2 
\nonumber\\
=& \sum_{j=1}^d[ n_j(1+m_j)+m_j(1+n_j)]
\label{qfimultimode}
\end{align}
where we defined the average photon numbers $n_j\equiv \bra{\phi_A} a_j^{\dagger}a_j\ket{\phi_A}$ and $m_j\equiv \bra{\phi_B} b_j^{\dagger}b_j\ket{\phi_B}$. 
In the particular case $n_i=m_i$, e.g. for twin states, we finally obtain
\begin{align}
\mathcal{Q}=  2\sum_{j=1}^d n_j (n_j+1) .
\label{qfimultimodeII} 
\end{align}
From this expression one can immediately recover the QFI of twin Fock states  $\mathcal{Q}=N(1+N/2)$ \cite{holland93}, which corresponds to $d=1$ and $n_1=N/2$. For twin superradiant states, from our considerations of Sec.  \ref{MostRelevantModes} we have that $2n_1\approx 0.90N$, $2n_2\approx 0.08N$ and $2n_3\approx 0.02N$, from which we obtain $\mathcal{Q}\approx 0.41N^2 + N$, hence recovering our previous results \cite{Paulisch2019}. 

More generally,   \eqref{qfimultimode} provides a simple and clear expression for the potential of a particular multimode state for optical interferometry, and from it we learn that
\begin{enumerate}
\item To obtain the QFI of a twin multimode state with $N/2$  photons in each arm, it is enough to compute the average photon number of the internal modes of each arm.
\item When the multimode state is generated through a non-linear decay as described in Section \ref{non-linear}, then our techniques enable us to compute the QFI for large photon number ($N$ up to $N \approx 1000$).
\item  Heinseberg scaling (i.e. $\mathcal{Q} \propto N^2$) is possible when the number of relevant modes is independent of $N$, and quantum-enhanced scaling (i.e. $\mathcal{Q}\propto N^{1+\alpha}$ with $\alpha>0$) when the number of relevant modes grows  sublinearly with $N$. 
\end{enumerate}

\subsection{Number resolved measurements and QFI}

Although the QFI provides the maximal sensitivity of $\ket{\psi_{\varphi}}$ to $\varphi$, it is equally important to understand how to achieve it in practice. For that, in this section we consider number-resolved measurements in both outputs of the interferometer. In particular, we consider two types of measurements
\begin{itemize}
\item Mode-Number-Resolved (MNR) measurements. That is, photon measurements that are able to distinguish both the specific mode and the number of photons. In this case, defining $P_{\alpha \beta}^{\rm (MNR)}(\varphi)\equiv P(\{\alpha_j\}_{j=1}^d,\{\beta_j\}_{j=1}^d|\varphi)$ and with \eqref{defabket} we have 
\begin{align}
\hspace{5mm}P_{\alpha \beta}^{\rm (MNR)}(\varphi)= \left| \langle \psi_{\varphi}|( |  \{\alpha_j\}\rangle_a \otimes |\{\beta_j\} \rangle_b )\right|^2.
\end{align} 
\item Number-Resolved (NR) measurements. We also consider standard photon counting measurements that are not able to distinguish between the different modes. In this case the corresponding probability distribution to obtain $n$ and $m$ photons in each output of the interferometer reads
\begin{align}
\hspace{5mm}P_{n m }^{\rm (NR)}(\varphi)=\sum_{\{\alpha\} \in \mathcal{C}(n),  \{\beta\} \in \mathcal{C}(m)}P_{\alpha, \beta}^{\rm (MNR)}(\varphi).
\label{NRmeas}\end{align} 
\end{itemize}
In both cases, we assume that the detector frequency-bandwidth is larger than the photonic wavepackets linewidth, which in the superradiant case scale as $\gamma \propto N/\ln N$ (see Sec. \ref{Sec:AvPhotNum} and Fig.~\ref{avPhoton}). 

Let us now compute the CFI~\eqref{cfi} for MNR and NR measurements when $\varphi \rightarrow 0$ by extending the considerations of~\cite{Pezze2013} to multimode states. We first expand  $P_{\alpha \beta}(\varphi)$ around $\varphi=0$:
\begin{align}
\hspace{-1mm}P_{\alpha, \beta}^{\rm (MNR)}(\varphi)\hspace{-0.5mm}=\hspace{-0.5mm}P_{\alpha, \beta}^{\rm (MNR)}(0)+ \dot{P}^{\rm (MNR)}_{\alpha \beta}(0)\varphi + \frac{1}{2} \ddot{P}^{\rm (MNR)}_{\alpha \beta}(0) \varphi^2 \hspace{-0.5mm}
\end{align}
up to order $\mathcal{O}(\varphi^3)$. 
Let us now consider this expansion for different cases:
\begin{enumerate}
\item $n$ photons in one output and $m$ photons in the other one, i.e., $P_{\alpha \beta}^{\rm (MNR)}(\varphi)$'s such that $\{\alpha_j\}\in \mathcal{C}(n)$ and $\{\beta_j\}\in \mathcal{C}(m)$.  Then using \eqref{derab} we  obtain, 
$P_{\alpha \beta}^{\rm (MNR)}(\varphi)=1+\mathcal{O}(\varphi^2),$
which does not contribute to \eqref{cfi}.  
\item $n\pm 1$ photons in one output and $m\mp 1$ photons in the other one.  That is, $P_{\alpha \beta}^{\rm (MNR)}(\varphi)$'s such that $\{\alpha_j\}\in \mathcal{C}(n\pm 1)$ and $\{\beta_j\}\in \mathcal{C}(m\mp 1)$.  Then again using \eqref{derab} we have
$P_{\alpha \beta}(\varphi)^{\rm (MNR)}=\frac{1}{2} \ddot{P}_{\alpha \beta}(0) \varphi^2$ with $\ddot{P}_{\alpha \beta}(0)=2|\braket{\dot{\psi}_{\varphi=0}}{\{\alpha_j\}_{j=1}^{d}, \{\beta_j\}_{j=1}^d}|^2$.   This case  does  contribute to \eqref{cfi}.
\item $n\pm m$ photons in one output and $m\mp m$ photons in the other one, with $m\geq 2$. Then we have,
$P_{\alpha \beta}(\varphi)=\mathcal{O}(\varphi^{m+1})$,
which again does not contribute to \eqref{cfi}.
\end{enumerate}
Putting together these considerations we obtain:
\begin{align}
\lim_{\varphi\rightarrow 0}\mathcal{C}&=  2 \sum_{\{\alpha_j\}\in \mathcal{C}(n\pm1), \{\beta_j \}\in \mathcal{C}(m\mp1)} \ddot{P}_{\alpha \beta}(0)
\label{derqfi}
\nonumber\\
&=  4\hspace{-1mm} \sum_{\{\alpha_j\}\in \mathcal{C}(n\pm1), \{\beta_j \}\in \mathcal{C}(m\mp1)} \hspace{-2mm} |\braket{\dot{\psi}_{\varphi=0}}{\{\alpha_j\}_{j=1}^{d}, \{\beta_j\}_{j=1}^d}|^2
\nonumber\\
&=4 \bra{\dot{\psi}_{\varphi=0}} \hspace{-1mm}\left(\hspace{-1mm}\sum_{\{\alpha_j\}, \{\beta_j \}}\hspace{-2mm} \ket{\{\alpha_j\}, \{\beta_j \}}\bra{\{\alpha_j\}, \{\beta_j \}} \right)\hspace{-1mm}\ket{\dot{\psi}_{\varphi=0}}
\nonumber\\ 
&=4   \braket{\dot{\psi}_{\varphi=0}}{\dot{\psi}_{\varphi=0}}
\nonumber\\ 
&=\lim_{\varphi\rightarrow 0} \mathcal{Q}.
\end{align}
where in the third line we used that $\ket{\dot{\psi}_{\varphi=0}}$ has only support in the subspace of $n\pm1$ photons in one arm and $m\mp 1$ in the other one, and in the fourth line we used $\braket{\psi_{\varphi=0}}{\dot{\psi}_{\varphi=0}}=0$.
Hence, we conclude that $\mathcal{C}= \mathcal{Q}$ around $\varphi=0$ for MNR measurements. 

 \begin{figure}
	\centering
	\includegraphics[width=1\linewidth]{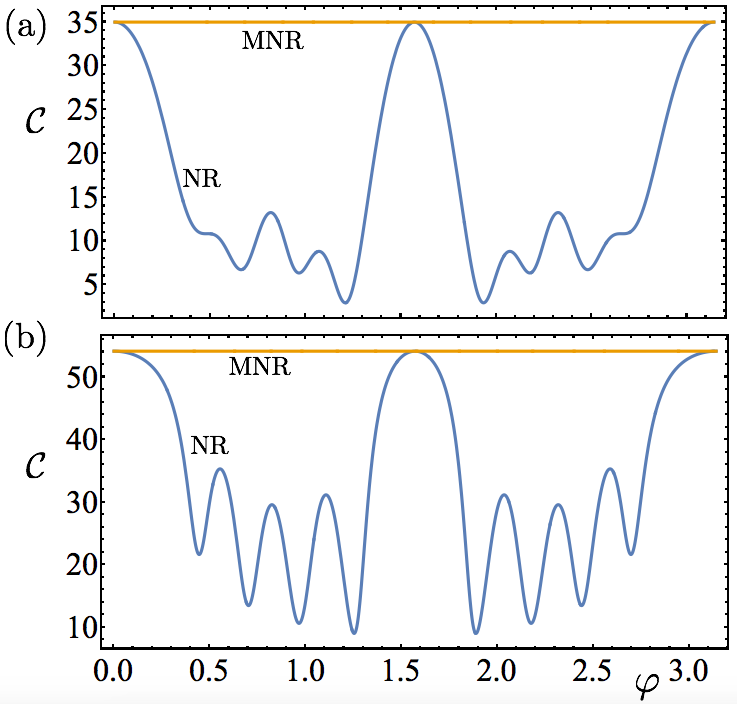}
	\caption{Classical Fisher information for MNR measurements (in orange) and for NR measurements (in blue) when two copies of $\ket{\psi_1}$ in \eqref{psiex}  (case (a)) or  two copies of $\ket{\psi_2}$ in \eqref{psiexII} (case (b)) enter through a MZ interferometer. The CFI for MNR measurements coincides with the QFI for all $\varphi$. }
	\label{numbresI}
\end{figure}

Crucially, the derivation  \eqref{derqfi} follows analogously for NR measurements, i.e., for the coarse-grained  distribution \eqref{NRmeas}. This has the very important consequence for practical implementations that experimentally one does not need to distinguish between the different modes to saturate the QFI. 
These conclusions hold around $\varphi=0$, but there is in principle no reason why they should also hold for other $\varphi$. To address this point, we consider two illustrative   states:
\begin{align}
\label{psiex}
&\ket{\phi_1}= \sum_{j=0}^n \frac{1}{\sqrt{n}} \ket{j,n-j}
\\
\label{psiexII}
&\ket{\phi_2}= \frac{1}{\sqrt{2}}\left(\ket{0,n}+\ket{n,0} \right),
\end{align}
and compute the CFI for the corresponding twin states  (i.e.  $\ket{\psi_1,\psi_1}$ or $\ket{\psi_2,\psi_2}$ are the input states of the MZ interferometer). 
In Figure \ref{numbresI} we plot the CFI for \eqref{psiex} and  \eqref{psiexII} with $n=5$ and given MNR and NR photon measurements. We observe that $\mathcal{C}=\mathcal{Q}$ for MNR measurements  whereas for the NR measurement  the equality is only saturated around $\varphi=0$ (or multiples of $\pi/2$).   It is worth stressing that one can always add phase shifters to compensate for $\varphi\neq 0$ during the estimation process in order to guarantee that $\mathcal{C}=\mathcal{Q}$ for NR measurements. These are crucial considerations  to take into account in implementations of quantum metrology with multimode states. It is also worth noticing that the multimode structure of the state leads to a non-trivial change of the CFI for NR measurements, as illustrated in Figure \ref{numbresI}.

\begin{figure}
	\centering
	\includegraphics[width=1\linewidth]{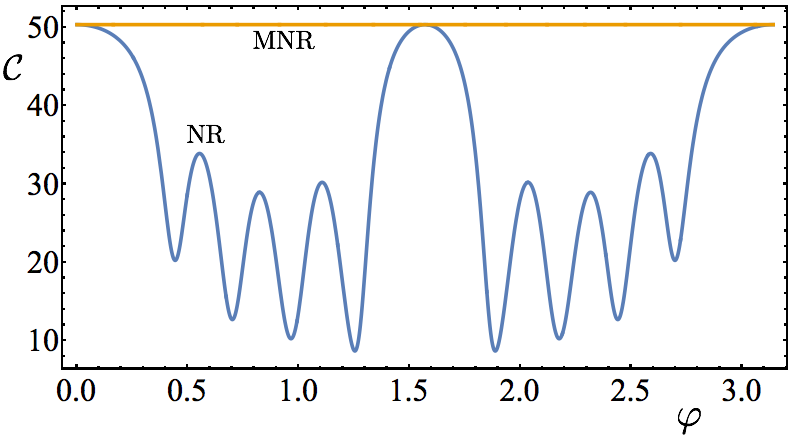}
	\caption{Classical Fisher information for MNR measurements (in orange) and for NR measurements (in blue) for a twin superradiant state with $N=10$ (i.e. $n=5$). 
	 }
	\label{numbresss}
\end{figure}

Finally,  we also compare NR and MNR measurements  for twin-superradiant states as shown  in Figure \ref{numbresss}, which also illustrates the optimality of choosing $\varphi \approx 0$. To obtain the results of Figure \ref{numbresss}, we have used the effective descriptions obtained in Section \ref{EffectiveDescriptions} which enable us to describe superradiant states through effective descriptions using a low number of modes (here we use two-mode descriptions, $d=2$, for simplicity). 

 While our techniques enable us to compute the metrological properties up to hundreds of photons, we note that current experimental implementations of photon-number detectors are limited to few photons resolution ($N\leq 10$), see e.g.~\cite{Hadfield2009,PhysRevA.99.043822}. It is also worth mentioning that recent theoretical proposals show that  atoms coupled to the waveguide could also be used for photon detection up to considerably larger photon numbers~\cite{Romero2009,Peropadre2011,malz2019number}, which hints to the exciting possibility that atoms suitably coupled to  waveguides can be both used to generate and detect photons. 
In the next section we discuss a proposal for quantum metrology with superradiant states in the presence of photon loss for realistic photon numbers, namely~$N\leq 10$.

\subsection{Photon loss in the interferometer and the measurement device}

Besides finding specific measurement schemes to saturate $\mathcal{Q}$, in practice it is also crucial to consider imperfections in the interferometer and in the measurement devices. In fact,   quantum-enhancements in metrology are largely affected by imperfections (either in the interferometer or due to imperfect measurements), as photon loss prevents Heisenberg scaling for sufficiently large $N$   \cite{fujiwara2008fibre,Knysh2011,escher2011general,janekdecoherence}. To intuitively understand why Heisenberg scaling is lost, note that for obtaining $\mathcal{C}=\mathcal{Q}$ in \eqref{derqfi}  requires photon-measurement detectors that are capable of distinguishing a single photon (i.e. between $N$ photons and $N \pm 1$ photons in the outcomes of the MZ interfereometer); yet, in the presence of any finite photon loss  (see Fig. \ref{MZscheme}), this is no longer possible when   $\eta N \simeq 1$, where  $\eta$ is the probability of losing a photon. Still, quantum enhancements in the presence of photon loss can  appear as a better prefactor in  $\mathcal{Q} \propto N$ \cite{Knysh2011,escher2011general,janekdecoherence} as  classical schemes are limited by the shot-noise limit $\mathcal{Q}\leq N$. This advantage however highly depends on the state into consideration: for example, GHZ states, which are optimal in ideal conditions, are known to quickly lose any sensitivity to $\varphi$ in the presence photon loss. Twin Fock States (TFS), on the other hand, are known to be a good candidate for quantum metrology even in the presence of photon loss $\eta$ (quantifying the probability of losing a photon in each arm of the interferometer), as in this case the QFI becomes 
\begin{align}
\mathcal{Q}^{\rm TFS}  = \frac{N}{2}\frac{1-\eta}{\eta}, \hspace{8mm} {\rm for} \hspace{2mm} \eta N \gg 1,
\label{scalingnoise}
\end{align}
which is half of the optimal one in the limit of large $N$~\cite{Knysh2011,knysh2014true}. It is also important to stress that although these considerations (e.g.  \eqref{scalingnoise}) are derived in the asymptotic limit ($\eta N \gg 1$),  they provide valuable insights already for moderate $N$    \cite{Knysh2011,knysh2014true}.  To extend \eqref{scalingnoise} and the general considerations of \cite{fujiwara2008fibre,Knysh2011,escher2011general,janekdecoherence} to multimode states is certainly an interesting but also  challenging endeavour, as the multimode structure of the state makes it difficult to diagonalise it to be able to compute~$\mathcal{Q}$. In this Section, we instead pursue a  more humble goal: We compare (twin) superradiant and Fock states for some specific case-studies of MZ interferometery  and find that they perform similarly even in the presence of photon loss (which is expected as superradiant states contain $0.9N$ photons in a single mode). 


We consider photon loss by adding a beam splitter with transmitivity $\eta$ before each measurement apparatus (this is equivalent to placing the beam splitters before the second beam splitter of the MZ interferometer as the losses are symmetric). This is implemented by adding orthogonal modes $\{e_j\}_{j=1}^D$, $\{f_j\}_{j=1}^D$ and implementing the transformations $\tilde{U}_{BS}$:
 \begin{align}
\begin{pmatrix} 
a_j  \\
e_j   
\end{pmatrix}
\xrightarrow{\tilde{U}_{BS}}
\frac{1}{\sqrt{2}}\begin{pmatrix} 
\sqrt{\eta} & \sqrt{1-\eta} \\
-\sqrt{1-\eta} & \sqrt{\eta}  
\end{pmatrix}
\begin{pmatrix} 
a_j  \\
e_j   
\end{pmatrix}
\end{align}
and
 \begin{align}
\begin{pmatrix} 
b_j  \\
f_j   
\end{pmatrix}
\xrightarrow{\tilde{U}_{BS}}
\frac{1}{\sqrt{2}}\begin{pmatrix} 
\sqrt{\eta} & \sqrt{1-\eta} \\
-\sqrt{1-\eta} & \sqrt{\eta}  
\end{pmatrix}
\begin{pmatrix} 
b_j  \\
f_j   
\end{pmatrix}.
\end{align}

We again characterise the input superradiant states through the effective descriptions obtained in Section \ref{EffectiveDescriptions} with $d=2$. These effective descriptions enable us to easily account for photon loss, which would be highly challenging through the continuous descriptions \eqref{genstate}. Still, our results are limited to low $N$, both because we can only obtain the coefficients of the state for $N<10$ (see \ref{EffectiveDescriptions}), and also because  the possible outcomes of the experiment (i.e. the size of the probability distribution $P_{\alpha, \beta}^{\rm (MNR)}(\varphi)$) grows as $\mathcal{O}(N^4)$ when dealing with two-mode states in each arm of the interferometer, hence making it difficult to compute the analytical expression \eqref{cfi} for high $N$.

\begin{figure}
	\centering
	\includegraphics[width=1\linewidth]{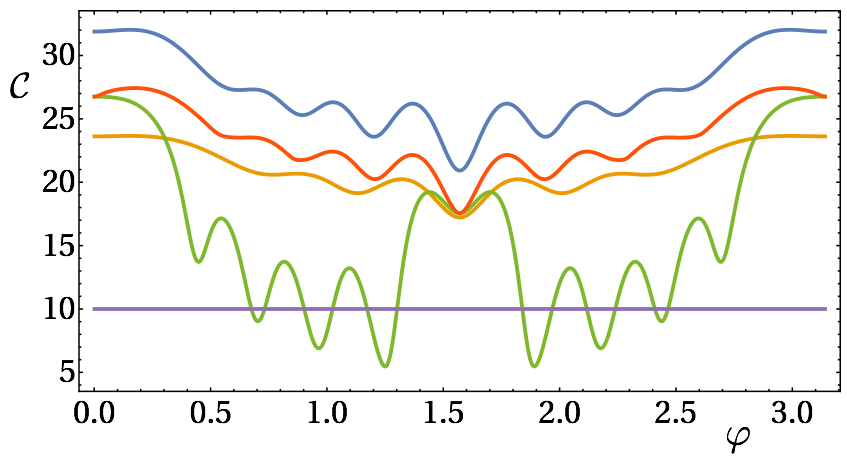}
	\caption{Classical Fisher information  in the presence of photon loss ($\eta=0.9$) for:  NR measurements in twin  Fock states of $N=8$ (orange) and $N=10$ (blue) photons, and twin superradiant states with $N=10$ and NR (green) and NMR (red) measurements. Finally, the straight purple line shows the classical Shot Noise Limit  (SNL) corresponding to $\mathcal{C}=N$. }
	\label{numbres}
\end{figure}

In Fig.~\ref{numbres} we show $\mathcal{C}$ in the presence of photon loss in the measurement devices for different configurations: twin  Fock states of $N=8$ (orange) and $N=10$ (blue) photons and NR measurements, and twin superradiant states with $N=10$ and NR (green) and NMR (red) measurements. We observe how twin superradiant states perform close to Fock states, and how NR measurements for superradiant (and hence multimode) states become optimal around $\varphi=0$, as expected from our previous considerations. It is worth pointing out that as $\eta$ increases, the optimal value moves away from $\varphi=0$, which happens for NR and NMR measurements, and  for single-mode and multimode states.  These observations are confirmed by further numerical results for other values of  $N$ and $\eta$, which are shown in Appendix \ref{NumPhotonLoss}. From these results, we conclude that twin-superradiant states behave fairly similar  to twin-Fock states in terms of their metrological performance also in the presence of photon loss; this conclusion is indeed expected given  that superradiant states contain $\approx 0.9N$ photons in a single mode. 

\section{Conclusions}
\label{conclusions}

To sum up, we have developed the theoretical tools to characterize (i.e., compute observables) of a wide class of multimode photonic states coming from the emission of a general non-linear level structure. Besides, we provide a constructive way of capturing the properties of these multimodal states with few-mode descriptions. To illustrate the potential of these tools, we have applied them to the case of superradiant photonic states, showing, for example, their observables can be captured efficiently already with two or three modes up to a large number of photons. Finally, we applied these ideas to a phase estimation proposal based on twin superradiant states and number resolved measurements. Our results suggest that twin superradiant states of $N$ photons  are a promising candidate for quantum metrology, as they perform approximately as a twin Fock state with $\approx 0.9 N$ photons, which are in fact the number of photons contained in a single mode. The crucial difference is that twin superradiant states can be  generated in a \emph{deterministic} and \emph{scalable} manner (assuming no photon loss and perfect control on the system's Hamiltonian to ensure the collective decay), in contrast to standard probabilistic methods to create Fock states, whose success probability decays exponentially with $N$ even in ideal conditions.  We hope these ideas motivate its experimental implementation in nanophotonic waveguides coupled to atoms~\cite{vetsch10aaa,thompson13a,goban13a, corzo16a,sorensen16a,solano17a}, artificial emitters~\cite{laucht12a,lodahl15a,sipahigil16a}, or molecules~\cite{faez14a}.

Our results in quantum optical interferometry are in fact general and can be applied to arbitrary multimode states with a fixed photon number.  Indeed, 
we have considered phase estimation in a Mach-Zender interferometer and  derived
a simple expression for the quantum Fisher information obtained when  the input state of each arm of the interferometer is a generic $n$-photon $d$-mode state, and shown that it can be saturated by  number-resolved measurements that  \emph{cannot}  distinguish between the different modes. This is a crucial observation for experiments, as it shows that single-mode proposals for quantum metrology, e.g. \cite{holland93,campos03a,Hofmann2009,Knysh2011,knysh2014true,Pezze2013,Oszmaniec2016}, can be naturally extended to the multimode regime without requiring extra resources in terms of the measurement devices.


\section{Acknowlegdements}

We thank Jan Ko\l{}ody\'{n}ski for insightful discussions. We acknowledge funding from ERC Advanced Grant QENOCOBA under the EU Horizon 2020  program (Grant Agreement No. 742102). AGT acknowledges funding from project PGC2018-094792-B-I00  (MCIU/AEI/FEDER, UE), CSIC Research Platform PTI-001, and CAM/FEDER Project No.~S2018/TCS-4342~(QUITEMAD-CM).

\newpage

\setcounter{equation}{0}
\setcounter{section}{0}
\makeatletter


\begin{widetext}
	\vspace{5mm}
		\hspace{7.5cm}\textbf{\large APPENDIX}
	\vspace{-1mm}	
		
\section{Recurrence relations}
\label{RecurrenceRelations}
Here we build recurrence relations to compute 
\begin{align}
f(\{x_j,y_j,\tilde{x}_j,\tilde{y}_j\}_{j=1}^{n})=\bra{\phi^{(N)}}b_{x_1,y_1} ... b_{x_n,y_n} b_{\tilde{x}_1,\tilde{y}_1}^{\dagger} ... b_{\tilde{x}_1,\tilde{y}_1}^{\dagger}\ket{\phi^{(N)}}.
\label{generalexpressionf}
\end{align}
As the derivation is rather non-trivial, for clarity we will start by computing simple but relevant cases where $n=0,1,2$ before deriving a general relation for \eqref{generalexpressionf}. 

Along the derivation we will use that from  \eqref{comm} it follows that
\begin{align}
\bra{0} b_{x_1,y_1} b^{\dagger}_{\tilde{x}_1,\tilde{y}_1} \ket{0}=\frac{2\sqrt{x_1 \tilde{x}_1}}{x_1 +\tilde{x}_1+2i(\tilde{y}_1-y_1)}
\label{prodxy}
\end{align}
and similarly
\begin{align}
[b_{x_1,y_1} ,b^{\dagger}_{\tilde{x}_1,\tilde{y}_1}]=\frac{2\sqrt{x_1 \tilde{x}_1}}{x_1 +\tilde{x}_1+2i(\tilde{y}_1-y_1)}.
\label{commxy}
\end{align}

\subsection{Normalisation}
It is instructive to first check   that $\ket{\phi^{(N)}}$ is normalised. We want to compute,
\begin{align}
\braket{\phi^{(N)}}{\phi^{(N)}}=\frac{1}{(N!)^2}\multInt_0^{\infty} A_{t_1 ... t_N } A_{s_1 ... s_N}^*  \left(\prod_{j,k=1}^N {\rm d} t_j \hspace{1mm}  {\rm d} s_j \right) \bra{0} a_{s_1}... a_{s_N}   a^{\dagger}_{t_1}... a^{\dagger}_{t_N}\ket{0}
\end{align}
Using the symmetry of $A_{t_1 ... t_N }$ under permutations over $\{ t_j\}$ and the commutation relation \eqref{comm}, we arrive at 
\begin{align}
\braket{\phi^{(N)}}{\phi^{(N)}}=\frac{1}{N!}\multInt_0^{\infty}\left(\prod_{j,k=1}^N {\rm d} t_j \right)  |A_{t_1 ... t_N}|^2 . 
\end{align}
In order to solve this integral, which includes a time-ordering operation $\mathcal{T}$, we  split the integral as a sum of integrals using  
\begin{align}
\int_0^{\infty} \hspace{-1mm}\int_0^{\infty} {\rm d} x \hspace{1mm} {\rm d} y \quad \mathcal{T}\langle \mathcal{O}_x \mathcal{O}_y \rangle =\int_0^{\infty} {\rm d} y\int_y^{\infty} {\rm d} x  \quad \langle \mathcal{O}_x \mathcal{O}_y \rangle +\int_0^{\infty} {\rm d} x\int_x^{\infty} {\rm d} y  \quad \langle \mathcal{O}_y \mathcal{O}_x \rangle.
\label{splittimeord}
\end{align}
There are $N!$ such integrals, and they are equivalent. Hence we have that
\begin{align}
\braket{\phi^{(N)}}{\phi^{(N)}}=\int_0^{\infty} {\rm d} t_N \hspace{1mm}...\int_{t_3}^{\infty} {\rm d} t_2 \hspace{1mm} \int_{t_2}^{\infty}  {\rm d} t_1 \hspace{1mm} \prod_{j=1}^N \gamma_j  \exp\left[(\gamma_{j-1}-\gamma_j)t_j\right]. 
\end{align}
This integral can be easily worked out using $\int_s^\infty {\rm d}s e^{-at}=e^{-as}/a $, which leads to the desired result
\begin{align}
\braket{\phi^{(N)}}{\phi^{(N)}}=1. 
\end{align}

\subsection{Average photon number}
We first consider $f(x_1,y_1,\tilde{x}_1,\tilde{y}_1)$. Before proceeding to its calculation, first note
\begin{align}
f(x_1,y_1,\tilde{x}_1,\tilde{y}_1)=\frac{2\sqrt{x_1 \tilde{x}_1}}{x_1 +\tilde{x}_1+2i(\tilde{y}_1-y_1)}+ \bra{\phi^{(N)}}b_{\tilde{x}_1,\tilde{y}_1}^{\dagger} b_{x_1,y_1} \ket{\phi^{(N)}}.
\end{align}
Using \eqref{comm} and \eqref{prodxy} and the symmetry of $A_{t_1 ... t_N}$ over permutations, we first have
\begin{align}
f(x_1,y_1,\tilde{x}_1,\tilde{y}_1)=\frac{2\sqrt{x_1 \tilde{x}_1}}{x_1 +\tilde{x}_1+2i(\tilde{y}_1-y_1)}+N I^{(N)}_1(x_1,y_1,x_2,y_2)
\end{align}
with
\begin{align}
I^{(N)}_1(x_1,y_1,\tilde{x}_1,\tilde{y}_1)=\frac{1}{N!}\int_0^{\infty} \hspace{-2mm}...\int_0^{\infty}\prod_{j=2}^N {\rm d} t_j \hspace{1mm} {\rm d} u\hspace{1mm} {\rm d} v \hspace{1mm} A_{u t_2 ... t_N }A_{v t_2 ... t_N }^* B_u^{(x_1,y_1)*} B_v^{(\tilde{x}_1,\tilde{y}_1)}.
\label{I_1}
\end{align}
 This integral is challenging to compute because of the time-ordering $\mathcal{T}$. We will compute the integral through a recurrence relation. The rough idea is as follows: first one splits the integral as a sum of integrals using \eqref{splittimeord}. Since the number of integrals increases exponentially with $N$, it is crucial to use that when any of the $t_j$'s is integrated, the resulting integral is the same due to the symmetry of $A$ under permutations. This allows us to keep the computation efficient, as the number of integrals grows linearly with $N$. Let us implement this idea by developing a recurrence relation where each step corresponds to an integration through one of the variables  of integration ($u,v, t_1, t_2...$). Let us start by defining the integrals 
\begin{align}
&F_{1,1}^{(N-j)}=\int_0^{\infty} \hspace{-2mm}...\int_0^{\infty} \left(\prod_{k=j+1}^N {\rm d} t_k \hspace{-1mm}\right){\rm d} u\hspace{1mm} {\rm d} v \hspace{1mm}  e^{-c_{1,1}^{(N-j)}\max\{t_j,u,v\}}\mathcal{T} \bra{\varphi_j}\mathcal{O}_{u}\mathcal{O}_{t_{j+1}}...\mathcal{O}_{t_N}\ket{\varphi_N} ( \bra{\varphi_j}\mathcal{O}_{v}\mathcal{O}_{t_{j+1}}...\mathcal{O}_{t_N}\ket{\varphi_N})^* B_u^{(x_1,y_1)*} B_v^{(\tilde{x}_1,\tilde{y_1})}
\nonumber\\
&F_{1,0}^{(N-j)}=\int_0^{\infty} \hspace{-2mm}...\int_0^{\infty} \left(\prod_{k=j+1}^N {\rm d} t_k\hspace{-1mm} \right) {\rm d} u\hspace{1mm} {\rm d} v \hspace{1mm}  e^{-c_{1,0}^{(N-j)}\max\{t_j,u\}}\mathcal{T} \bra{\varphi_j}\mathcal{O}_{u}\mathcal{O}_{t_{j+1}}...\mathcal{O}_{t_N}\ket{\varphi_N} ( \bra{\varphi_j}\mathcal{O}_{t_{j+1}}...\mathcal{O}_{t_N}\ket{\varphi_N})^* B_u^{(x_1,y_1)*} 
\nonumber\\
&F_{0,1}^{(N-j)}=\int_0^{\infty} \hspace{-2mm}...\int_0^{\infty} \left(\prod_{k=j+1}^N {\rm d} t_k \hspace{-1mm}\right)  {\rm d} u\hspace{1mm} {\rm d} v \hspace{1mm}  e^{-c_{0,1}^{(N-j)}\max\{t_j,v\}}\mathcal{T} \bra{\varphi_j}\mathcal{O}_{t_{j+1}}...\mathcal{O}_{t_N}\ket{\varphi_N} ( \bra{\varphi_j}\mathcal{O}_{v}\mathcal{O}_{t_{j+1}}...\mathcal{O}_{t_N}\ket{\varphi_N})^* B_v^{(\tilde{x}_1,\tilde{y}_1)}
\nonumber\\
&F_{0,0}^{(N-j)}=\int_0^{\infty} \hspace{-2mm}...\int_0^{\infty} \left(\prod_{k=j+1}^N {\rm d} t_k \hspace{-1mm}\right) {\rm d} u\hspace{1mm} {\rm d} v \hspace{1mm}  e^{-c_{0,0}^{(N-j)}\max\{t_j\}}\mathcal{T} \bra{\varphi_j}\mathcal{O}_{t_{j+1}}...\mathcal{O}_{t_N}\ket{\varphi_N} ( \bra{\varphi_j}\mathcal{O}_{t_{j+1}}...\mathcal{O}_{t_N}\ket{\varphi_N})^* 
\end{align}
with
\begin{align}
&c_{1,1}^{(N-j)}=\gamma_{j-1}
\nonumber\\
&c_{1,0}^{(N-j)}=\frac{1}{2}\tilde{x}_1+\frac{1}{2}(\gamma_{j-1}+\gamma_{j})-i(\tilde{y}_1-(\omega_{j-1}-\omega_j))
\nonumber\\
&c_{0,1}^{(N-j)}=\frac{1}{2}x_1+\frac{1}{2}(\gamma_{j-1}+\gamma_{j})+i(y_1-(\omega_{j-1}-\omega_j))
\nonumber\\
&c_{0,0}^{(N-j)}=\frac{1}{2}(x_1+\tilde{x}_1)+\gamma_{j}+i(y_1-\tilde{y}_1),
\end{align}
with $j=1,...,N$. 
Note that 
\begin{align}
I^{(N)}_1(x_1,y_1,\tilde{x}_1,\tilde{y}_1)=\frac{1}{N!}F^{(N-1)}_{1,1}.
\label{fxy}
\end{align}
We can then compute $F^{(N-1)}_{1,1}$ by noting the following recurrence relation (with $j=1,...,N$):
\begin{align}
&F^{(N-j)}_{1,1}=\frac{(N-j) \gamma_j}{c_{1,1}^{(N-j-1)}}F^{(N-j-1)}_{1,1}+\frac{ \sqrt{\tilde{x}_1\gamma_j}}{c_{1,0}^{(N-j)}}F^{(N-j)}_{1,0}+\frac{ \sqrt{x_1\gamma_j}}{c_{0,1}^{(N-j)}}F^{(N-j)}_{0,1}
\nonumber\\
&F^{(N-j)}_{1,0}=\frac{(N-j) \sqrt{\gamma_j\gamma_{j+1}}}{c_{0,1}^{(N-j-1)}}F^{(N-j-1)}_{1,0}+\frac{ \sqrt{x_1\gamma_j}}{c_{0,0}^{(N-j)}}F^{(N-j)}_{0,0}
\nonumber\\
&F^{(N-j)}_{0,1}=\frac{(N-j) \sqrt{\gamma_j\gamma_{j+1}}}{c_{1,0}^{(N-j-1)}}F^{(N-j-1)}_{0,1}+\frac{ \sqrt{\tilde{x}_1\gamma_j}}{c_{0,0}^{(N-j)}}F^{(N-j)}_{0,0}
\nonumber\\
&F^{(N-j)}_{0,0}=\frac{(N-j) \gamma_{j+1}}{c_{0,0}^{(N-j-1)}}F^{(N-j-1)}_{0,0},
\label{recurrencerelation}
\end{align}
together with
\begin{align}
F^{(0)}_{0,0}=1.
\end{align}
Next, the idea is to express \eqref{recurrencerelation} as a matrix multiplication. In particular, let us define a matrix of size $(4N)^2$ given by:
\[M=
\begin{bmatrix}
    M_{11}       & M_{12} & M_{13} & M_{14} \\
    M_{21}       & M_{22} & M_{23}  & M_{24}\\
    M_{31}       & M_{32} & M_{33}  & M_{34} \\
    M_{41}       & M_{42} & M_{43}  & M_{44}
\end{bmatrix}
\]
where $M_{ij}$ are matrices of size $N^2$ with entries $M_{ij}[[k,l]]$ given by
\begin{align}
&M_{11}[[j+1,j]]=j\frac{\gamma_{N-j+1}}{c_{0,0}^{(j-1)}} \hspace{2mm} {\rm with} \hspace{2mm} j=1,...N-1, \hspace{5mm} {\rm and} \hspace{2mm} 0 \hspace{2mm} {\rm otherwise}
\nonumber\\
&M_{22}[[j+1,j]]=j\frac{\sqrt{\gamma_{N-j}\gamma_{N-j+1}}}{c_{1,0}^{(j-1)}}\hspace{2mm} {\rm with} \hspace{2mm} j=1,...N-1, \hspace{5mm} {\rm and} \hspace{2mm} 0 \hspace{2mm} {\rm otherwise}
\nonumber\\
&M_{33}[[j+1,j]]=j\frac{\sqrt{\gamma_{N-j}\gamma_{N-j+1}}}{c_{0,1}^{(j-1)}}\hspace{2mm} {\rm with} \hspace{2mm} j=1,...N-1, \hspace{5mm} {\rm and} \hspace{2mm} 0 \hspace{2mm} {\rm otherwise}
\nonumber\\
&M_{44}[[j+1,j]]=j\frac{\gamma_{N-j}}{c_{1,1}^{(j-1)}} \hspace{2mm} {\rm with} \hspace{2mm} j=1,...N-1, \hspace{5mm} {\rm and} \hspace{2mm} 0 \hspace{2mm} {\rm otherwise}
\nonumber\\
&M_{21}[[j,j]]=\frac{\sqrt{\gamma_{N-j}x}}{c_{0,0}^{(j)}} \hspace{2mm} {\rm with} \hspace{2mm} j=1,...N,\hspace{5mm} {\rm and} \hspace{2mm} 0 \hspace{2mm} {\rm otherwise}
\nonumber\\
&M_{31}[[j,j]]=\frac{\sqrt{\gamma_{N-j}y}}{c_{0,0}^{(j)}} \hspace{2mm} {\rm with} \hspace{2mm} j=1,...N,\hspace{5mm}{\rm and} \hspace{2mm} 0 \hspace{2mm} {\rm otherwise}
\nonumber\\
&M_{42}[[j,j]]=\frac{\sqrt{\gamma_{N-j}y}}{c_{1,0}^{(j)}} \hspace{2mm} {\rm with} \hspace{2mm} j=1,...N,\hspace{5mm}{\rm and} \hspace{2mm} 0 \hspace{2mm} {\rm otherwise}
\nonumber\\
&M_{43}[[j,j]]=\frac{\sqrt{\gamma_{N-j}x}}{c_{0,1}^{(j)}} \hspace{2mm} {\rm with} \hspace{2mm} j=1,...N, \hspace{5mm}{\rm and} \hspace{2mm} 0 \hspace{2mm} {\rm otherwise}
\end{align} 
and the remaining submatrices are zero. Defining the initial vector: $\vec{v}_0=\{1,0,...,0\}$, one finds that
\begin{align}
\vec{v}_f=M^{N+1}.\vec{v}_0
\end{align}
where 
\begin{align}
\vec{v}_f=\{0,...,0,F^{(N-1)}_{1,1}\},
\end{align}
which provides the desired result. When computing this numerically, it is convenient to compute instead
\begin{align}
\vec{v}_f=\left(\prod_{j=1}^N \frac{M}{j}\right).M.\vec{v}_0
\end{align}
which directly provides $I^{(N)}_1(x_1,y_1,\tilde{x}_1,\tilde{y}_1)$ in \eqref{fxy}.

\subsection{Two-photon correlators}
Let us now move to the computation of
\begin{align}
f(\{x_j,y_j,\tilde{x}_j,\tilde{y}_j\}_{j=1}^{2})=\bra{\phi^{(N)}}b_{x_1,y_1} b_{x_2,y_2} b_{\tilde{x}_1,\tilde{y}_1}^{\dagger}  b_{\tilde{x}_2,\tilde{y}_2}^{\dagger}\ket{\phi^{(N)}}. 
\end{align}
Using \eqref{comm} and \eqref{prodxy} and the symmetry of $A_{t_1 ... t_N}$ over permutations, we arrive at
\begin{align}
f(\{x_j,y_j,\tilde{x}_j,\tilde{y}_j\}_{j=1}^{2})=&\frac{2\sqrt{x_1 \tilde{x}_1y_1 \tilde{y}_1}}{\left(x_1 +\tilde{x}_1+2i(\tilde{y}_1-y_1)\right)\left( x_2 +\tilde{x}_2+2i(\tilde{y}_2-y_2)\right)}+\frac{2\sqrt{x_1 \tilde{x}_1y_1 \tilde{y}_1}}{\left(x_1 +\tilde{x}_2+2i(\tilde{y}_2-y_1)\right)\left( x_2 +\tilde{x}_1+2i(\tilde{y}_1-y_2)\right)}
\nonumber\\
&+N \frac{2\sqrt{x_1 \tilde{x}_1}}{x_1 +\tilde{x}_1+2i(\tilde{y}_1-y_1)} I_1^{(N)}(x_2,y_2,\tilde{x}_2,\tilde{y}_2)+N \frac{2\sqrt{x_1 \tilde{x}_2}}{x_1 +\tilde{x}_2+2i(\tilde{y}_2-y_1)} I_1^{(N)}(x_2,y_2,\tilde{x}_1,\tilde{y}_1)
\nonumber\\
&+N \frac{2\sqrt{x_2 \tilde{x}_1}}{x_2 +\tilde{x}_1+2i(\tilde{y}_1-y_2)} I_1^{(N)}(x_1,y_1,\tilde{x}_2,\tilde{y}_2)+N \frac{2\sqrt{x_2 \tilde{x}_2}}{x_2 +\tilde{x}_2+2i(\tilde{y}_2-y_2)} I_1^{(N)}(x_1,y_1,\tilde{x}_1,\tilde{y}_1)
\nonumber\\ 
&+N(N-1)I^{(N)}_2(x_1,y_1,x_2,y_2,\tilde{x}_1,\tilde{y}_1,\tilde{x}_2,\tilde{y}_2)
\end{align}
where we have defined
\begin{align}
I^{(N)}_2(x_1,y_1,&x_2,y_2,\tilde{x}_1,\tilde{y}_1,\tilde{x}_2,\tilde{y}_2)=
\nonumber\\ 
&\frac{1}{N!} \int_0^{\infty} \hspace{-2mm}...\int_0^{\infty} \left( \prod_{j=3}^N {\rm d} t_j \right)\hspace{1mm} {\rm d} u_1\hspace{1mm}{\rm d} u_2\hspace{1mm}{\rm d} v_1\hspace{1mm} {\rm d} v_2 \hspace{1mm} A_{u_1 u_2 t_3 ... t_N }A_{v_1 v_2 t_3 ... t_N }^* B_{u_1}^{(x_1,y_1)*} B_{u_2}^{(x_2,y_2)*}B_{v_1}^{(\tilde{x}_1,\tilde{y}_1)}B_{v_2}^{(\tilde{x}_2,\tilde{y}_2)}
\end{align}
in analogy with \eqref{I_1}. Identifying 
\begin{align}
I^{(N)}_2(x_1,y_1,&x_2,y_2,\tilde{x}_1,\tilde{y}_1,\tilde{x}_2,\tilde{y}_2)=\frac{F^{(N-2)}_{1,1,1,1}}{N!}
\end{align}
we find the following recurrence relation (a natural extension of \eqref{recurrencerelation}),
\begin{align}
F^{(N-1-k)}_{1-s_1,1-s_2,1-t_1,1-t_2}=&\delta_{s_1,0}F^{(N-1-k)}_{0,1-s_2,1-t_1,1-t_2} \frac{\sqrt{x_1 \gamma_{k+s_2}}}{c^{(N-1-k)}_{0,1-s_2,1-t_1,1-t_2}}+\delta_{s_2,0}F^{(N-1-k)}_{1-s_1,0,1-t_1,1-t_2} \frac{\sqrt{x_2 \gamma_{k+s_1}}}{c^{(N-1-k)}_{0,1-s_2,1-t_1,1-t_2}}
\nonumber\\
&+\delta_{t_1,0}F^{(N-1-k)}_{1-s_1,1-s_2,0,1-t_2} \frac{\sqrt{\tilde{x}_1 \gamma_{k+t_2}}}{c^{(N-1-k)}_{1-s_1,1-s_2,0,1-t_2}}+\delta_{t_2,0}F^{(N-1-k)}_{1-s_1,1-s_2,1-t_1,0} \frac{\sqrt{\tilde{x}_2 \gamma_{k+t_1}}}{c^{(N-1-k)}_{1-s_1,1-s_2,1-t_1,0}}
\nonumber\\
&+(N-k-1)F^{(N-2-k)}_{1-s_1,1-s_2,1-t_1,1-t_2} \frac{\sqrt{\gamma_{k+s_1+s_2} \gamma_{k+t_1+t_2}}}{c^{(N-2-k)}_{1-s_1,1-s_2,1-t_1,1-t_2}}
\label{recrelvar}
\end{align}
with $k=1,...,N-2$, and 
\begin{align}
F^{(0)}_{0,0,0,0}=1,
\end{align}
and 
\begin{align}
c^{(N-2-k)}_{1-s_1,1-s_2,1-t_1,1-t_2}=\sum_{i=1}^2 s_i \left(\frac{x_i}{2}+iy_i \right) +t_i \left(\frac{\tilde{x}_i}{2} -i\tilde{y}_i \right) + \frac{1}{2}\left(\gamma_{s_1+s_2+k} + \gamma_{t_1+t_2+k}\right)+i(\omega_{k+t_1+t_2}-\omega_{k+s_1+s_2}).
\end{align}
In order to compute the recurrence relation \eqref{recrelvar} as a matrix multiplication, it convenient to define basis vectors: $\ket{k,s_1,s_2,t_1,t_2} $
with $k=\{1,...,N-1\}$, and $s_1,s_2,t_1,t_2 \in \{0,1\}$.  Then, the idea is to define a matrix $M$ that satisfies,
\begin{align}
M \ket{k,s_1,s_2,t_1,t_2}&= C_{k-1,s_1,s_2,t_1,t_2}\ket{k-1,s_1,s_2,t_1,t_2}+C_{k,s_1-1,s_2,t_1,t_2}\ket{k,s_1-1,s_2,t_1,t_2}
\nonumber\\&+C_{k,s_1,s_2-1,t_1,t_2}\ket{k,s_1,s_2-1,t_1,t_2}+C_{k,s_1,s_2,t_1-1,t_2}\ket{k,s_1,s_2,t_1-1,t_2}
\nonumber\\&+C_{k,s_1,s_2,t_1,t_2-1}\ket{k,s_1,s_2,t_1,t_2-1}
\end{align}
with
\begin{align}
&C_{k-1,s_1,s_2,t_1,t_2}=(N-k-1)\frac{\sqrt{\gamma_{k+s_1+s_2} \gamma_{k+t_1+t_2}}}{c^{(N-2-k)}_{1-s_1,1-s_2,1-t_1,1-t_2}}
\nonumber\\
&C_{k,s_1-1,s_2,t_1,t_2}=\delta_{s_1,0} \frac{\sqrt{x_1 \gamma_{k+s_2}}}{c^{(N-1-k)}_{0,1-s_2,1-t_1,1-t_2}}
\nonumber\\
&C_{k,s_1,s_2-1,t_1,t_2}=\delta_{s_2,0} \frac{\sqrt{x_2 \gamma_{k+s_1}}}{c^{(N-1-k)}_{0,1-s_2,1-t_1,1-t_2}}
\nonumber\\
&C_{k,s_1,s_2,t_1-1,t_2}=\delta_{t_1,0} \frac{\sqrt{y_1 \gamma_{k+t_2}}}{c^{(N-1-k)}_{1-s_1,1-s_2,0,1-t_2}}
\nonumber\\
&C_{k,s_1,s_2,t_1,t_2-1}=\delta_{t_2,0} \frac{\sqrt{y_2 \gamma_{k+t_1}}}{c^{(N-1-k)}_{1-s_1,1-s_2,1-t_1,0}}
\end{align}
which provide the coefficients of  $M$. Then, notice that from \eqref{recrelvar} one obtains,
\begin{align}
M^{N+2}|N-1,1,1,1,1\rangle = F^{(N-2)}_{1,1,1,1}\ket{1,0,0,0,0},
\end{align}
which gives the desired result. Note that the matrices $M$ are now of size $(2^4(N-1))^2$.


\subsection{Higher order terms}
\label{sec:higherorder}

Given these previous considerations, it is in principle not difficult (but quite tedious) to extend these techniques to higher-order correlators of the form $\bra{\phi^{(N)}}b_{x_1,y_1} ... b_{x_n,y_n} b_{\tilde{x}_1,\tilde{y}_1}^{\dagger} ... b_{\tilde{x}_n,\tilde{y}_n}^{\dagger}\ket{\phi^{(N)}}$. Essentially, following the previous considerations we need to compute integrals of the form 
\begin{align}
I^{(N)}_n(\{x_j,y_j,\tilde{x}_j,\tilde{y}_j\}_{j=1}^{n})=\frac{1}{N!} \int_0^{\infty} \hspace{-2mm}...\int_0^{\infty}\left( \prod_{j=n+1}^N {\rm d} t_j \right) \hspace{1mm}\left(\prod_{j=1}^n {\rm d} u_j\hspace{1mm} {\rm d} v_j\hspace{1mm}  B_{u_j}^{(x_j,y_j)*}B_{v_j}^{(\tilde{x}_j,\tilde{y}_j)} \right)&A_{u_1 ...u_n t_{n+1} ... t_N }A_{v_1... v_n t_{n+1} ... t_N }^* 
\end{align}
In complete analogy with the previous considerations, we can define
\begin{align}
I^{(N)}_n(\{x_j,y_j,\tilde{x}_j,\tilde{y}_j\}_{j=1}^{n})=\frac{F^{(N-n)}_{1,...,1}}{N!}
\end{align}
and the following recurrence relation can be derived (a natural extension of \eqref{recrelvar}),
\begin{align}
F^{(N-n+1-k)}_{\{1-s_i\}_{i=1}^n,\{1-t_i\}_{i_1}^n}=& \sum_{j=1}^n \delta_{s_j,0}F^{(N-n+1-k)}_{\{1-s_i\}_{i=1}^{j-1},0,\{1-s_i\}_{i=j+1}^n,\{1-t_i\}_{i_1}^n}  \frac{\sqrt{x_j \gamma_{k+\sum_{j=1}^ns_j}}}{c^{(N-n+1-k)}_{\{1-s_i\}_{i=1}^j,0,\{1-s_i\}_{i=j+1}^n,\{1-t_i\}_{i_1}^n}}
\nonumber\\
&+ \sum_{j=1}^n \delta_{t_j,0}F^{(N-n+1-k)}_{\{1-s_i\}_{i=1}^n,\{1-t_i\}_{i=1}^{j-1},0,\{1-t_i\}_{i=j+1}^n}  \frac{\sqrt{\tilde{x}_j \gamma_{k+\sum_{j=1}^n t_j}}}{c^{(N-n+1-k)}_{\{1-s_i\}_{i=1}^j,0,\{1-s_i\}_{i=j+1}^n,\{1-t_i\}_{i_1}^n}}
\nonumber\\
&+(N-k-n+1)F^{(N-k-n)}_{\{1-s_i\}_{i=1}^n,\{1-t_i\}_{i_1}^n} \frac{\sqrt{\gamma_{k+s_1+s_2} \gamma_{k+t_1+t_2}}}{c^{(N-k-n)}_{\{1-s_i\}_{i=1}^n,\{1-t_i\}_{i_1}^n}}
\label{recrelvarII}
\end{align}
with $k=1,...,N-n$ and 
\begin{align}
F^{(0)}_{0,...,0}=1,
\end{align}
and
\begin{align}
c^{(N-k-n+1)}_{1-s_1,1-s_2,1-t_1,1-t_2}=\sum_{i=1}^n s_i \left(\frac{x_i}{2}+iy_i \right) +t_i \left(\frac{\tilde{x}_i}{2} -i\tilde{y}_i \right)  + \frac{1}{2}\left(\gamma_{\sum_{i=1}^n s_i+k} + \gamma_{\sum_{i=1}^n t_i+k}\right)+i(\omega_{k+\sum_{i=1}^n t_i}-\omega_{k+\sum_{i=1}^n s_i}).
\label{generalc}
\end{align}
This recurrence relation \eqref{recrelvarII} can be computed through  a matrix multiplication of $M^{\mathcal{O}(N)}$ in analogy with the previous sections, where $M$ is now a matrix of size  $(2^{2n} (N-n+1))^2$. Hence we notice that the complexity of the calculation grows exponentially with the order of the correlator. 

\subsection{Overlap}
 \begin{figure}
	\centering
	\includegraphics[width=0.6\linewidth]{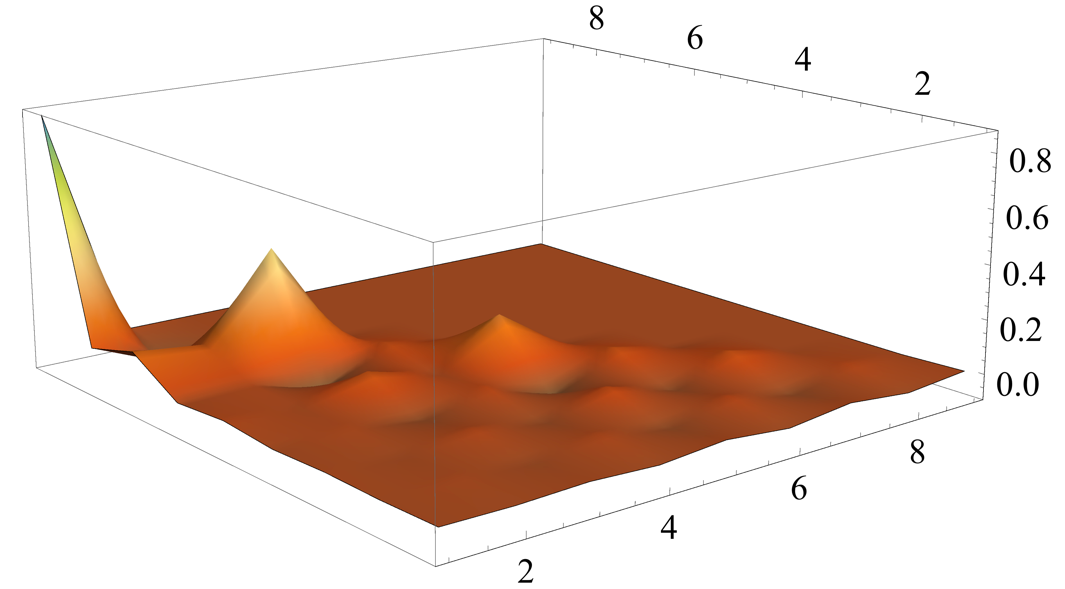}
	\caption{Coefficients of the three-mode approximation of a superradiant state for $N=8$ and $D=8$.}
	\label{coefIII}
\end{figure}
Consider the computation of $\langle 0 | c_1^{k_1}c_2^{k_2}... c_d^{k_d}  \ket{\phi^{(N)}}$ with $\sum_{i=1}^d k_i=N$.  This can expressed by a linear combination of products of the form 
\begin{align}
\label{appendixcoeff}
\langle 0 | b_{x_1}^{k_1}b_{x_2}^{k_2}... b_{x_D}^{k_D}  \ket{\phi^{(N)}}= \int_0^{\infty}\hspace{-2mm}...\int_0^{\infty} \left(\prod_{j=1}^N \hspace{-0.5mm} {\rm d} t_j \right) \left( \hspace{2mm}  \prod_{j=1}^{k_1} B_{t_j}^{(x_1)*}\right) \left( \prod_{j=k_1+1}^{k_1+k_2} B_{t_j}^{(x_2)*} \right)...\left(\prod_{j=N-k_D+1}^{N} B_{t_j}^{(x_D)*}\right)  A_{t_1...t_N}
\end{align}
 where we used \eqref{defd_k}. As in the previous section, we can compute each integral by solving the following recurrence relation:
\begin{align}
F_{\{k_i-s_i\}_{i=1}^D}=\sum_{j=1}^D \theta[k_j-s_j] (k_j-s_j) F_{\{k_i-s_i\}_{i=1}^{j-1},k_j-s_j-1,\{k_i-s_i\}_{i=j+1}^{D}}\frac{\sqrt{x_j \gamma_{\sum_k s_k}}}{c_{\{k_i-s_i\}_{i=1}^{j-1},k_j-s_j-1,\{k_i-s_i\}_{i=j+1}^{D}}}
\end{align}
where $\theta[x]$ is the step function ($\theta[-|x|]=0$ and $\theta[|x|]=1$),
together with the initial condition
\begin{align}
F_{\{0\}_{j=1}^D=1}=1,
\end{align}
and the coefficients
\begin{align}
c_{\{k_i-s_i\}_{i=1}^{D}}=\frac{1}{2}\left( \sum_{j=1}^D s_j x_j+\gamma_{\sum_j s_j} \right). 
\end{align}
This provides the desired solution as:
\begin{align}
F_{\{k_i\}_{i=1}^D}=\langle 0 | b_{x_1}^{k_1}b_{x_2}^{k_2}... b_{x_D}^{k_D}  \ket{\phi^{(N)}}
\end{align}
Following the same logic as in the following sections, this integral can be computed by a product of matrices in a space of dimension 
\begin{align}
{\rm dim}=\prod_{j=1}^D k_j,
\end{align} 
which is approximately bounded as
\begin{align}
N \lesssim {\rm dim} \lesssim \left( N/D \right)^D .
\end{align}

In Figure \ref{coefIII} we illustrate these ideas by computing all $\langle 0 | c_1^{k_1}c_2^{k_2} c_3^{k_3}  \ket{\phi^{(N)}}$ with $N=8$, $D=8$ and $d=3$.

\section{Numerical results on quantum metrology with photon loss}
\label{NumPhotonLoss}
This section shows more numerical results on the CFI with photon loss and considering as input states twin Fock states and twin superradiant states, as shown in Figures \ref{numbresII} ,\ref{numbresIII} and \ref{numbresIV}. 

 \begin{figure}
	\centering
	\includegraphics[width=0.49\linewidth]{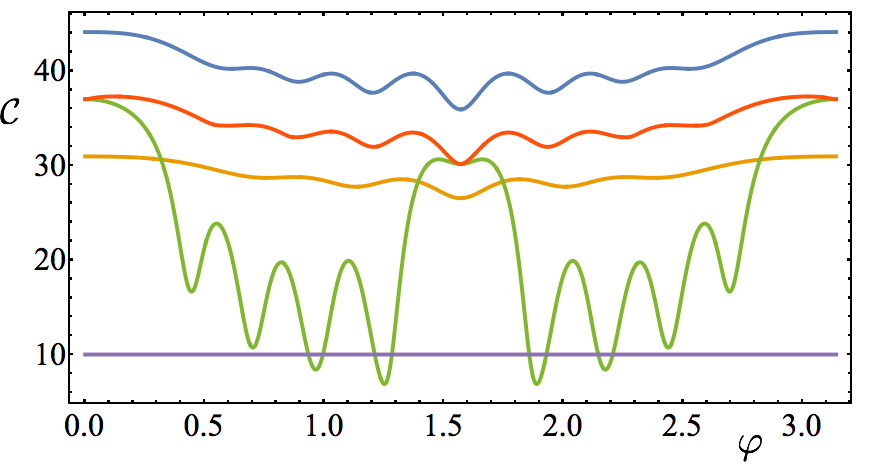}
	\caption{Classical Fisher information  in the presence of photon loss ($\eta=0.95$) for:  NR measurements in twin  Fock states of $N=8$ (orange) and $N=10$ (blue) photons, and twin superradiant states with $N=10$ and NR (green) and NMR (red) measurements. Finally, the straight purple line shows the classical Shot Noise Limit  (SNL) corresponding to $\mathcal{C}=N$. }
	\label{numbresII}
\end{figure}
\begin{figure}
	\centering
	\includegraphics[width=0.49\linewidth]{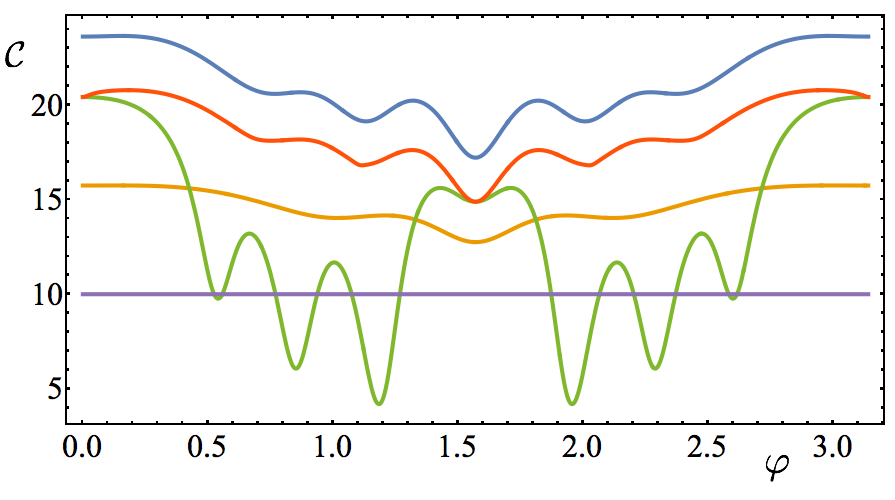}
	\caption{Classical Fisher information  in the presence of photon loss ($\eta=0.9$) for:  NR measurements in twin  Fock states of $N=6$ (orange) and $N=8$ (blue) photons, and twin superradiant states with $N=8$ and NR (green) and NMR (red) measurements. }
	\label{numbresIII}
\end{figure}
 \begin{figure}
	\centering
	\includegraphics[width=0.49\linewidth]{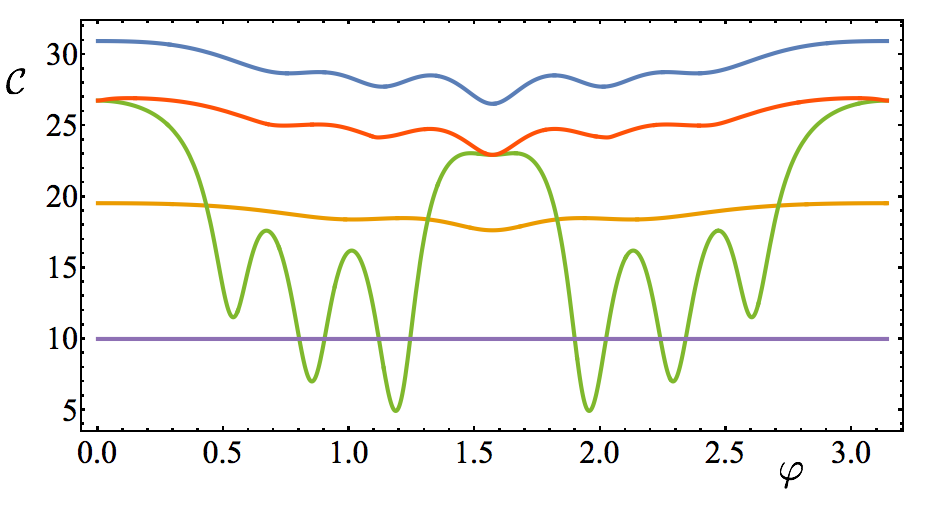}
	\caption{Classical Fisher information  in the presence of photon loss ($\eta=0.95$) for:  NR measurements in twin  Fock states of $N=8$ (orange) and $N=10$ (blue) photons, and twin superradiant states with $N=8$ and NR (green) and NMR (red) measurements. }
	\label{numbresIV}
\end{figure}


%
%

\end{widetext}

\end{document}